# On some many-valued abstract logics and their $\in_T$-style extensions

Steffen Lewitzka*

September 24, 2018

**Abstract**

Logical systems with classical negation and means for sentential or propositional self-reference involve, in some way, paradoxical statements such as the liar. However, the paradox disappears if one replaces classical by an appropriate non-classical negation such as a paraconsistent one (no paradox arises if the liar is both true and false). We consider a non-Fregean logic which is a revised and extended version (Lewitzka 2012) of $\in_T$-Logic (epsilon-T-Logic) originally introduced by (Sträter 1992) as a logic with a total truth predicate and propositional quantifiers. Self-reference is achieved by means of equations between formulas which are interpreted over a model-theoretic universe of propositions. Paradoxical statements, such as the liar, can be asserted only by unsatisfiable equations and do not correlate with propositions. In this paper, we generalize $\in_T$-Logic to a four-valued logic related to Dunn/Belnap logic $B_4$. We also define three-valued versions related to Kleene's logic $K_3$ and Priest's Logic of Paradox $P_3$, respectively. In this many-valued setting, models may contain liars and other "paradoxical" propositions which are ruled out by the more restrictive classical semantics. We introduce these many-valued non-Fregean logics as extensions of abstract parameter logics such that parameter logic and extension are of the same logical type. For this purpose, we define and study abstract logics of type $B_4$, $K_3$ and $P_3$. Using semantic methods we show compactness of the consequence relation of abstract logics of type $B_4$, give a representation as minimally generated logics and establish a connection to the approach of (Font 1997). Finally, we present a complete sequent calculus for the $\in_T$-style extension of classical abstract logics simplifying constructions originally developed by (Sträter 1992, Zeitz 2000, Lewitzka 1998).

*Universidade Federal da Bahia - UFBA, Instituto de Matemática, Departamento de Ciência da Computação, Salvador da Bahia – BA, Brazil, e-mail: steffen@dcc.ufba.br

# 1 Introduction

This article represents a thorough revision and extension of work originally developed in the unpublished papers [13] and [14].

$\in_T$-Logic was designed by Werner Sträter [25] in the early 1990's as a theory of truth and propositional self-reference. The project was part of a larger research program on self-referential structures and non-classical set theory supervised by Prof. Bernd Mahr at TU Berlin. $\in_T$-Logic contains the classical propositional connectives, a connective for propositional identity as well as propositional quantifiers and operators (in postfix notation) that represent a total truth predicate in Tarski's sense: $\varphi : true$ reads "$\varphi$ is true" and $\varphi : false$ reads "$\varphi$ is false". Formulas are interpreted over a model-theoretic universe of propositions. This universe is divided into two disjoint subsets $TRUE$ and $FALSE$ — the sets of the true and the false propositions, respectively. The identity connective plays the crucial role for expressing propositional (self-) reference. A formula $\varphi \equiv \psi$ expresses that $\varphi$ and $\psi$ denote the same proposition of the given model-theoretic universe. Since there is no distinction between formulas and terms, equations such as $c \equiv (c : true)$, $c \equiv (c : false)$ and $x \equiv (x \to \varphi)$ can be formulated. The first formula asserts that the proposition denoted by the constant symbol $c$ is the proposition "$c$ is true". Thus, the first equation says that $c$ denotes a truth teller. Similarly, the second equation asserts that $c$ denotes a liar proposition, and the third equation says that variable $x$ denotes a contingent liar. The first equation is satisfiable, i.e. there are models containing truth tellers. The classical truth conditions, however, ensure that the second equation is unsatisfiable and thus a contradictory formula of $\in_T$-Logic. Although the liar can be asserted by that equation there is no liar proposition in the universe of any model. The satisfaction of the third equation depends on the truth value of $\varphi$. A model satisfying that equation necessarily satisfies formula $\varphi$. Further self-referential statements involving truth, falsity and classical connectives can be asserted without restrictions by means of equations. Semantic antinomies, such as the liar, are paradoxical, i.e., asserted by contradictory equations and are consequently ruled out by the classical truth conditions of the model-theoretic semantics. This is essentially the (rather elegant) solution to the semantic paradoxes proposed by Sträter [25]. The well known Tarski biconditionals (Tarski's T-scheme) can be expressed in the object language: $\varphi : true \leftrightarrow \varphi$, for every formula $\varphi$. In fact, the truth predicate on the object level coincides with the truth predicate of the metalanguage, which is given by model-theoretic satisfaction. Moreover, the Tarski biconditionals can be formulated via propositional quantifiers by a single theorem: $\forall x.(x : true \leftrightarrow x)$.

In the present paper, we abandon the classical setting and propose a many-valued semantics which involves more models. In particular, there will be models satisfying equations which are unsatisfiable in the classical setting. Solutions to such equations then will be self-referential propositions, such as the liar, which do not occur as propositions in the classical setting.[1] For this it is sufficient to abandon the constraint that the propositional universe $M$ of a model is the disjoint union of the sets $TRUE$ and $FALSE$. This involves the following 4 possibilities for every model-theoretic universe $M$:

(i) $M$ is the disjoint union of $TRUE$ and $FALSE$ (the classical case)

(ii) $M = TRUE \cup FALSE$ and $TRUE \cap FALSE \neq \varnothing$

(iii) $M \smallsetminus (TRUE \cup FALSE) \neq \varnothing$ and $TRUE \cap FALSE = \varnothing$

(iv) $M \smallsetminus (TRUE \cup FALSE) \neq \varnothing$ and $TRUE \cap FALSE \neq \varnothing$

Thus, every proposition $m \in M$ has exactly one of the following 4 possible truth values: $m$ is true and true only, $m$ is false and false only, $m$ is both true and false, $m$ is neither true nor false. This leads us to the well-known 4-valued propositional logic due to Dunn and Belnap which we denote here by $B_4$. The semantics of this logic is based on the above truth values which are denoted by $1, 0, B, N$, respectively. Considering the partial order $\leq$ defined by $0 \leq N$, $0 \leq B$, $N \leq 1$ and $B \leq 1$ one gets a well-known complete lattice of truth values with designated values $1$ and $B$. The sublattices of truth values $\{0, N, 1\}$ and $\{0, B, 1\}$ with designated values $1, N$ and $1, B$ correspond to the 3-valued Kleene logic $K_3$ and to the 3-valued paraconsistent logic $P_3$ of Priest, respectively. For an overview of these many-valued logics we refer the reader to [22], where Priest's logic $P_3$ is called $LP$, and the 4-valued logic of Dunn/Belnap is primarily discussed as the logic of First Degree Entailment.

As we saw above, an equation $\varphi \equiv \psi$ is interpreted in $\in_T$-Logic as "$\varphi$ and $\psi$ have the same denotation". Of course, one expects that $(\varphi \equiv \psi) \to (\varphi \leftrightarrow \psi)$ is a theorem. Since the denotation of a formula is not only given by a truth value but as an element of a model-theoretic universe, i.e. a proposition, the converse $(\varphi \leftrightarrow \psi) \to (\varphi \equiv \psi)$ is in general false. This latter implication was called by Roman Suszko the Fregean Axiom. Suszko proposed a program to develop logics without Fregean Axiom [26, 27]. $\in_T$-Logic can be seen as a non-Fregean logic in the sense of Suszko (although it was developed independently and under different assumptions aiming at a theory of truth and self-reference). In fact, the axioms of the Sentential Calculus with Identity SCI [4], which is the basic propositional non-Fregean logic, form a subsystem of a complete axiomatization of $\in_T$-Logic such as first given by Zeitz [29] and in slightly different and extended form by Lewitzka [19].[2] The SCI-axioms also derive from the system of axioms presented in [16], where a quantifier-free $\in_T$-style logic is studied. Given the axiomatic approach, $\in_T$-Logic can be characterized by the following equation: "$\in_T$-Logic = SCI + truth predicate + propositional quantifiers". From the semantic point of view, however, the connection between SCI and $\in_T$-Logic is less obvious since the respective models are defined in rather different ways.[3] These differences, however, are not essential and can be overcome. The Bloom-Suszko-style models of SCI are algebraic in the sense that the connective of the language correspond to operations on the model-theoretic universe. The interpretation of the language is managed by a semantic function which maps formulas to elements of the model-theoretic universe establishing an homomorphism between algebras of the same similarity type. On the other hand, the algebraic structure of Sträter's $\in_T$-style models is not explicitly given but is implicitly imposed by a semantic function that satisfies certain truth conditions and structural properties. Of crucial importance here is the substitution property. In [16] it is shown that in the quantifier-free context both styles of semantics are equivalent. Further discussion, with many historical details, can be found in [24] where also the relationship between $\in_T$-Logic and a further logic with propositional identity and propositional quantifiers, due to Hermes [10], is investigated.

---

[1] In some sense one may compare this situation with the extension of a field, say the field of reals, by new elements that satisfy originally unsatisfiable equations such as $x^2 + 1 = 0$.

In order to generalize the semantics of classical $\in_T$-Logic towards a 4-valued logic with the capability to extend abstract parameter logics we need an abstract approach to $B_4$. We will define a concept of "$B_4$ abstract logic" using similar methods as in [12, 13, 20]). This enables us to define an $\in_T$-style extension of a given $B_4$ abstract logic such that the extension turns out to be a $B_4$ abstract logic, too. First ideas to extend abstract logics by the expressive power of classical $\in_T$-Logic were developed by Zeitz [29].[4] Zeitz considers an abstract logic as a set of formulas together with a set of subsets of formulas, called the basis. The basis induces in the usual way a consequence relation. Motivated by topological concepts, Zeitz also defines certain mappings between abstract logics and proves some properties.[5]

---

[2] Sträter's original deductive system [25] is a sequent calculus.

[3] Indeed, connections to Suszko's non-Fregean logics remained unnoticed in the works of Zeitz and Sträter and were first discussed in [15].

[4] The $\in_T$-extension of a concrete classical first-order logic with a sound and complete sequent calculus, also for the extension, was presented in the author's Diplomarbeit [11].

[5] The approach was further developed in [12]. At that time it remained unnoticed by the author that similar concepts were already defined and investigated by Suszko, Bloom, Brown and others (see, e.g., [2, 3, 8]. Indeed, several abstract concepts of *logic* have been introduced independently in the literature over the last decades. For instance, much of the algebraic and topological machinery of van Fraassen's theory of *valuation spaces* [28] (see also [6] for a short overview) can be translated into the language of the theory of *abstract logics* in the sense of Suszko/Brown/Bloom, and vice-

However, Zeitz does not develop sufficient machinery and methods to distinguish between classical and certain non-classical abstract logics — any abstract logic is extended by *classical* $\in_T$-Logic in [29]. This may lead to combinations of non-classical connectives (coming from the parameter logic) with classical connectives (coming from $\in_T$-Logic) involving undesired and counter-intuitive phenomena.

Properties of connectives in abstract logics can be characterized by means of the corresponding consequence relation, i.e., closure operator (see, e.g., [3] for the classical case and [9] for the intuitionistic case) or by means of the minimal generator set (see [12, 13, 20, 17] for the classical, the intuitionistic and further non-classical cases). In [17] an intuitionistic $\in_T$-style extension of intuitionistic abstract logics is proposed. The truth conditions of the connectives of the underlying intuitionistic parameter logic are preserved in the extension which turns out to be an intuitionistic abstract logic, too. In the present article, we follow a similar strategy: First, we define a specific class of non-classical abstract logics by means of conditions over connectives and an appropriate generator set (we will work here with the set of *complete* theories and then show that the minimal generator set exists).[6] Then we define the $\in_T$-style extension in such a way that the truth conditions of the non-classical connectives remain preserved. That is, we define a non-classical $\in_T$-Logic which belongs to the same class of abstract logics as the underlying non-classical parameter logic. For this purpose, we must find four-valued interpretations for the connectives and quantifiers of $\in_T$-Logic. As in $\in_I$-Logic [15, 17], which is an intuitionistic and quantifier-free version of $\in_T$-Logic, we regard *truth* as non-classical satisfaction and we regard *falsity* as non-classical negation (and vice-versa). The non truth-functional connectives for identity and reference are interpreted classically in $\in_I$. That is, formulas of the form $\varphi \equiv \psi$ and $\varphi < \psi$ are either true or false at every world. This does not change the status of $\in_I$ as an intuitionistic abstract logic. However, if we want to construct a four-valued logic corresponding to $B_4$, then we cannot interpret such formulas classically: $\varphi \equiv \varphi$ and $c < (c : true)$ would be theorems and their negations would be unsatisfiable formulas. Recall that $B_4$ has neither theorems nor contradictory formulas. Fortunately, it is not hard to give a four-valued interpretation for these connectives. For a four-valued interpretation of quantifiers we are inspired by [21].

---

versa.

[6]The method to characterize certain non-classical abstract logics by means of minimal generator sets was developed in [13].

## 2  Some basic theory of Abstract Logics

Several abstract notions of logic were introduced and studied in the literature. One of the most accepted notions is the concept of Abstract Logic introduced by Brown, Suszko, Bloom [2, 4] as structures consisting of an algebra of formulas together with a closure operator or – equivalently – with a closure system on the set of formulas. The approach was further developed over the last decades within the research field of Abstract Algebraic Logic (see [8] for an overview). In the present article, we continue work started in [12, 13, 20] and study abstract logics as (topped or non-topped) intersection structures. An intersection structure can be seen as a closure system (or as a meet-semilattice). A major difference to the original approach due to Brown, Suszko, Bloom is that we consider minimally generated logics and characterize the properties of connectives by means of the minimal generator set instead of closure operators. This method developed from the study of Zeitz's [29] presentation of abstract logics and was fully described in [13] where intuitionistic and several other non-classical connectives are characterized by means of the minimal generator set of a certain intersection structure. The approach is picked up and further investigated in [20] where it is also shown that in the intuitionistic (and classical) case it is equivalent to the approach via closure operator [9]. In the following we will work essentially with definitions coming from [12, 13, 20].

**Definition 2.1** *An abstract logic $\mathcal{L} = (Expr_\mathcal{L}, Th_\mathcal{L}, \mathcal{C}_\mathcal{L})$ is given by a set $Expr_\mathcal{L}$ of formulas (or expressions), a subset $Th_\mathcal{L}$ of the power set of $Expr_\mathcal{L}$, called the set of $\mathcal{L}$-theories, and a set $\mathcal{C}_\mathcal{L}$ of operations on $Expr_\mathcal{L}$, called connectives. The following intersection axiom is satisfied: If $\mathcal{T} \subseteq Th_\mathcal{L}$ and $\mathcal{T} \neq \emptyset$, then $\bigcap \mathcal{T} \in Th_\mathcal{L}$.*

The set $\mathcal{C}_\mathcal{L}$ contains in general only those connectives which are *under consideration* and determine certain algebraic or topological properties. There may exist further operators and connectives in the language. For instance, one may consider a distributive abstract logic $\mathcal{L}$ with $\mathcal{C}_\mathcal{L} = \{\vee, \wedge\}$ which possibly contains also a classical or non-classical negation $\neg$. An (non-truth-functional) identity connective or reference connective of an (abstract) non-Fregean logic $\mathcal{L}$ is generally not considered as an element of $\mathcal{C}_\mathcal{L}$ since it does not involve interesting algebraic or topological properties.

An abstract logic $\mathcal{L}$ is an intersection structure in the sense of [5]. That is, the elements of $Th_\mathcal{L} \cup \{Expr_\mathcal{L}\}$ form a closure system. The corresponding closure operator $cl_\mathcal{L}: Pow(Expr_\mathcal{L}) \to Pow(Expr_\mathcal{L})$ gives rise to the consequence relation of $\mathcal{L}$: $A \Vdash_\mathcal{L} a :\Leftrightarrow a \in cl_\mathcal{L}(A)$. The intersection structure $\mathcal{L}$ is topped if $Expr_\mathcal{L} \in Th_\mathcal{L}$, otherwise $\mathcal{L}$ is non-topped. In the former case, we say that $\mathcal{L}$ is singular, in the latter case we say that $\mathcal{L}$ is regular. A set $A$ of formulas is $\mathcal{L}$-consistent if it is contained in some theory, otherwise $A$ is $\mathcal{L}$-inconsistent. Note that in a singular logic all sets of formulas are consistent, and in a regular logic a set $A$ is consistent iff there is a formula $b$ such that $A \nVdash_\mathcal{L} b$. The intersection axiom ensures that a set $T$ of formulas is a theory iff $T$ is consistent and deductively closed (under $\Vdash_\mathcal{L}$).

Depending on the particular situation and the point of view we may regard an abstract logic as a (topped or non-topped) intersection structure, a closure system, a meet-semilattice or a partial ordering – together with the given connectives.

Among the theories $Th_\mathcal{L}$ of a logic $\mathcal{L}$ we want to identify exactly those theories which can be seen as the theories of abstract models. We call such theories *stable* (with respect to the given connectives). Stable theories can be used to formulate the truth conditions of the given connectives in the usual model-theoretic fashion. It might be intuitively clear that the set of all theories which are not intersections of other theories should be stable. Guided by this intuition, the concept of a minimal generator set of a logic $\mathcal{L}$ is studied in [12, 13, 20]. It turns out that the minimal generator set is the set of all totally prime theories.

**Definition 2.2** *[[12, 13, 20]] We say that a theory $T \in Th_\mathcal{L}$ is generated by a set $\mathcal{G} \subseteq Th_\mathcal{L}$ if there is a non-empty $\mathcal{T} \subseteq \mathcal{G}$ such that $T = \bigcap \mathcal{T}$. $\mathcal{G}$ properly generates $T$ if there is a non-empty $\mathcal{T} \subseteq \mathcal{G}$ such that $T = \bigcap \mathcal{T}$ and $T \notin \mathcal{G}$.[7] A generator set of $\mathcal{L}$ is a subset $\mathcal{G} \subseteq Th_\mathcal{L}$ that generates all theories. $\mathcal{L}$ is called minimally generated if there is a minimal generator set $\mathcal{G}$ (minimal w.r.t. set-theoretic inclusion). Let $\kappa \geq \omega$ be a cardinal. A theory $T$ is $\kappa$-prime if $T = \bigcap \mathcal{T}$ implies $T \in \mathcal{T}$ for any non-empty $\mathcal{T} \subseteq Th_\mathcal{L}$ of cardinality $< \kappa$. A theory $T$ is called totally prime if $T$ is $\kappa$-prime for all infinite cardinals $\kappa$. We refer to $\omega$-prime theories simply as prime theories. A maximal theory is a theory which is maximal w.r.t. set-theoretic inclusion. We denote the sets of maximal, totally prime and prime theories by $MTh_\mathcal{L}$, $TPTh_\mathcal{L}$ and $PTh_\mathcal{L}$, respectively.*

The notions of *generator set* and *totally prime theory* are very similar to the respective order-theoretic notions of *meet-dense subset* of a complete lattice (in the context of closure spaces also called *basis*) and *completely meet-irreducible* or *completely meet-prime* element of a distributive, complete lattice. However, there is the following difference which justifies the use of our terminology in the context of abstract logics: In [20] it is shown that the $\kappa$-prime theories are precisely the theories stable w.r.t. $\kappa$-disjunction (see Theorem 2.7 below). In particular, if the logic is singular (a topped intersection structure), then the whole set of formulas is consistent and therefore a theory which is trivially stable w.r.t. $\kappa$-disjunction. Thus, the set of all formulas is $\kappa$-prime. This is in accordance with our definition of generator set and prime theory. However, it does not harmonize with the usual order-theoretic definitions where the whole set of formulas is not regarded as completely meet-prime and is not an element of the smallest meet-dense subset (the smallest basis). Therefore, it makes sense to use a new terminology here. Nevertheless, in the case of regular logics (non-topped intersection structures) our notions of totally prime theory and generator set coincide with the well-known order-theoretic concepts of completely prime element and meet-dense subset of a complete lattice, respectively. Also note that our notion of *theory* corresponds to the notion of (lattice-) filter on the set of formulas if and only if the logic has a connective for conjunction and the empty set is not a theory. There are logics where the empty set as well as the set of all formulas are prime theories (for instance, in $B_4$ abstract logics). By definition, these sets cannot be prime filters.

---

[7]The condition "$\mathcal{T} \neq \varnothing$" is missing in [Definition 2.2, [20]]. This condition is relevant in the case of singular logics: the theory of all formulas should not be properly generated.

Suppose $\mathcal{L}$ is minimally generated. If $T$ is $\kappa$-prime and $\kappa \geq \lambda \geq \omega$, then $T$ is $\lambda$ prime. It follows that a theory is prime iff it is $\kappa$-prime for some $\kappa \geq \omega$. A theory $T$ is totally prime iff $T$ is not the meet of any non-empty set of theories all distinct from $T$; in other words, $T$ is not properly generated by any set. It follows that $TPTh_\mathcal{L} \subseteq \mathcal{G}$ for any generator set $\mathcal{G}$. Moreover, if $\mathcal{G}$ is a minimal generator set, then $TPTh_\mathcal{L} = \mathcal{G}$. In order to see this, suppose $T \in \mathcal{G} \smallsetminus TPTh_\mathcal{L}$. Then $T = \bigcap \mathcal{T}$ for some non-empty set of theories $\mathcal{T}$ with $T \notin \mathcal{T}$. Each element of $\mathcal{T}$ is the meet of some non-empty subset of $\mathcal{G}$. We may assume that no one of these subsets contains $T$. It follows that $T$ is the meet of a non-empty subset of $\mathcal{G}$ and this subset does not contain $T$. Thus, $\mathcal{G} \smallsetminus \{T\}$ properly generates $T$ and is therefore a generator set, in contradiction to the minimality of $\mathcal{G}$. Thus, $TPTh_\mathcal{L}$ is the least generator set.

Furthermore, one easily checks that $MTh_\mathcal{L} \subseteq TPTh_\mathcal{L} \subseteq PTh_\mathcal{L}$ holds. Note that if $Expr_\mathcal{L} \in Th_\mathcal{L}$ (the logic is singular), then $Expr_\mathcal{L}$ is a totally prime theory, but it is not a completely meet-prime element in the meet-semilattice.

**Definition 2.3** *[[13, 20]] Let $\mathcal{L} = (Expr_\mathcal{L}, Th_\mathcal{L}, \{\curlywedge, \curlyvee, \sim, \rightarrowtail\})$ be a minimally generated abstract logic. $\mathcal{L}$ is an intuitionistic abstract logic if the following conditions are satisfied:*

(i) *The consequence relation is compact, that is, $A \Vdash_\mathcal{L} a$ implies the existence of a finite $A' \subseteq A$ with $A' \Vdash_\mathcal{L} a$.*

(ii) *For all expressions $a, b \in Expr_\mathcal{L}$ and for all $T \in TPTh_\mathcal{L}$ the following truth conditions of the connectives hold:*

- $a \curlywedge b \in T \iff a \in T$ and $b \in T$
- $a \curlyvee b \in T \iff a \in T$ or $b \in T$
- $\sim a \in T \iff T \cup \{a\}$ is inconsistent
- $a \rightarrowtail b \in T \iff$ for all $T' \in TPTh_\mathcal{L}$ with $T' \supseteq T$, if $a \in T'$ then $b \in T'$.

*An intuitionistic abstract logic $\mathcal{L}$ is a classical abstract logic if $MTh_\mathcal{L} = TPTh_\mathcal{L}$.*

The idea to specify an abstract logic by conditions on connectives w.r.t. the minimal generator set of the underlying intersection structure already appeared in [12], was fully developed in [13] and further studied in [20]. In [13], several non-classical connectives (such as intuitionistic, weak and paraconsistent negations) were defined by this method. Also the three-valued abstract logics of Kleene and Priest were defined there by means of conditions over a presupposed minimal generator set. Definition 2.3 above can also be found in similar form in [20] where additionally $\kappa$-conjunction and $\kappa$-disjunction ($\kappa \geq \omega$) are introduced. The condition that a logic is minimally generated seems to be rather natural. It is well-known that each of the equivalent conditions given in the following Fact is sufficient for the existence of a minimal generator set (see, e.g., Corollary 2.12 in [20]). The equivalences follow from well-known order-theoretic results (see, e.g., [5]).

**Fact 2.4** *For any abstract logic, the following conditions are equivalent:*

- *the consequence relation is compact*
- *the associated intersection structure is closed under unions of directed families (i.e., it is algebraic)*
- *the associated intersection structure is closed under unions of non-empty chains of theories*

*If one of these conditions is true, then the logic is minimally generated and the set of all totally prime theories is the smallest generator set.*

The next fact will be useful.

**Fact 2.5** *[Theorem 2.11, [20]] Let $\mathcal{T}$ be a generator set of logic $\mathcal{L}$. If the union of any non-empty chain of elements of $\mathcal{T}$ is an element of $\mathcal{T}$, then $\mathcal{L}$ is minimally generated, i.e. $TPTh_\mathcal{L}$ is the minimal generator set.*

Let $\mathcal{L}$ be any abstract logic. We saw in Definition 2.3 in which way the elements of the totally prime theories, can be used to define the intended truth conditions of the given connectives in a certain model-theoretic way. This works in intuitionistic and classical abstract logics and we expect that it works in other abstract logics, too. We call these conditions the *defining conditions of the connectives*. So the totally prime theories are stable by definition. Are there further sets of stable theories? This question can be formulated as follows. Are there further sets $\mathcal{T} \supseteq TPTh_\mathcal{L}$ of theories such that we get true conditions if we replace $TPTh_\mathcal{L}$ by $\mathcal{T}$ in the defining conditions of the connectives? A greatest stable set, if it exists, is called the set of complete theories, notation: $CTh_\mathcal{L}$ (see [13, 20]. This generalizes the notion of complete theory in classical logic, where the set of complete theories is given by the set of maximal theories, which is the minimal generator set (the greatest and the smallest stable set are here identical). In an intuitionistic abstract logic the greatest stable set is precisely the set of all $\omega$-prime theories (in the sense of Definition 2.2) as we will outline in the following.

Suppose $\mathcal{L}$ is an intuitionistic abstract logic. Recall that we call a theory $T$ ($\omega$-) prime if $T$ is not properly generated by any non-empty finite set of theories. That is, $T = T_1 \cap T_2$ implies $T = T_1$ or $T = T_2$, for any two theories $T_1, T_2$. In order-theory, such subsets of a lattice $T$ are called meet-irreducible (we may regard disjunction and conjunction as lattice operations). It is well-known (see, e.g., [23]) that in a distributive lattice a subset $T$ is meet-irreducible iff $T$ is a prime filter (i.e., $T$ is a proper filter and $a \vee b \in T$ implies $a \in T$ or $b \in T$). Since the theories of our intuitionistic abstract logic $\mathcal{L}$ can be seen as filters, it follows that the ($\omega$-) prime theories are precisely the prime theories in the usual sense of intuitionistic logic (i.e., theories $T$ with the property: $a \curlyvee b \in T$ iff $a \in T$ or $b \in T$). This fact also follows as a special case from our more general result Theorem 3.4 in [20]. Unfortunately, the main argument of our proof of Theorem 3.4 [20] is somewhat involved and contains a gap. We present here a revised and corrected proof (Theorem 2.7 below). For this we will need the following definition (see Definition 3.1 in [20]).

**Definition 2.6** *For some cardinal $\kappa \geq \omega$ let $\mathcal{L} = (Expr_\mathcal{L}, Th_\mathcal{L}, \mathcal{C})$ be a minimally generated abstract logic such that the connective $\bigvee_\kappa$ with the following defining condition belongs to $\mathcal{C}$. For all $a, b \in Expr_\mathcal{L}$, for all $A \subseteq Expr_\mathcal{L}$ with $|A| < \kappa$, and for all $T \in TPTh_\mathcal{L}$: $\bigvee_\kappa A \in T \Leftrightarrow A \cap T \neq \emptyset$. Then we say that $\mathcal{L}$ has $\kappa$-disjunction.*

**Theorem 2.7** *Let $\kappa \geq \omega$ be a cardinal and let $\mathcal{L} = (Expr_\mathcal{L}, Th_\mathcal{L}, \{\bigvee_\kappa\})$ be a minimally generated abstract logic with $\kappa$-disjunction $\bigvee_\kappa$. Then the greatest stable set, i.e. the set of complete theories $CTh_\mathcal{L}$, is exactly the set of all $\kappa$-prime* *theories.*[8] *In particular, $\kappa = \omega$ implies $CTh_\mathcal{L} = PTh_\mathcal{L}$.*

**Proof.** We must show that the set of theories satisfying the defining condition of the connective of $\kappa$-disjunction is exactly the set of $\kappa$-prime theories.

Suppose $T$ is not $\kappa$-prime. Then $T = \bigcap \mathcal{T}$ for some non-empty $\mathcal{T} \subseteq Th_\mathcal{L}$ with $T \notin \mathcal{T}$ and $|\mathcal{T}| < \kappa$. Let $\mathcal{T} = \{T_i \mid i < \beta\}$, $\beta < \kappa$. For each $i < \beta$ there is some $j < \beta$ such that $T_i \nsubseteq T_j$. Hence, for each $i < \beta$ we can choose a $j < \beta$ and an element $a_i \in T_i \smallsetminus T_j$. Let $\{a_i \mid i < \beta\}$ be the set of all elements chosen in this way. This set has cardinality $< \kappa$. Furthermore, $\bigvee_\kappa \{a_i \mid i < \beta\} \in T_j$ for all $j < \beta$. This follows from the fact that every $T_j$ is the intersection of a non-empty set of totally prime theories each of them containing $a_j$ and therefore also $\bigvee_\kappa \{a_j \mid j < \beta\}$, according to the defining condition of $\bigvee_\kappa$. Hence, $\bigvee_\kappa \{a_i \mid i < \beta\} \in T$. By construction, $\{a_i \mid i < \beta\} \cap T = \varnothing$. This contradicts the defining condition of $\bigvee_\kappa$. Thus, $T$ cannot be a complete theory. It follows that every complete theory is $\kappa$-prime.

Suppose now that $T$ is not a complete theory, $T \notin CTh_\mathcal{L}$. In particular, $T$ is not totally prime. There is some set $A \subseteq Expr_\mathcal{L}$ of cardinality $\mu < \kappa$ such that $\bigvee_\kappa A \in T$ and $A \cap T = \varnothing$. $T = \bigcap \mathcal{T}$ for some set $\mathcal{T}$ of totally prime theories. Note that $T \notin \mathcal{T}$ since $T$ is not totally prime. If $\mathcal{T}$ has cardinality less than $\kappa$, then $T$ is not $\kappa$-prime and we are done. So let us suppose $|\mathcal{T}| \geq \kappa$. The idea is to divide $\mathcal{T}$ into (at most) $\mu$-many suitable subsets. Let $A = \{a_i \mid i < \mu\}$. We have $\bigvee_\kappa A \in T'$ for each $T' \in \mathcal{T}$. Thus, each $T' \in \mathcal{T}$ contains some element of $A$. On the other hand, for each $a_i \in A$ there is some $T' \in \mathcal{T}$ such that $a_i \notin T'$. For $i, j < \mu$ we put $\mathcal{T}_{ij} = \{T' \in \mathcal{T} \mid a_i \notin T' \text{ and } a_j \in T'\}$. Note that the $\mathcal{T}_{ij}$ are *proper* subsets of $\mathcal{T}$ (a given $a_j$ cannot be contained in *all* $T' \in \mathcal{T}$).[9] Furthermore, we put $\mathcal{T}_\mu = \{T' \in \mathcal{T} \mid A \subseteq T'\}$.[10] Then $\mathcal{T} = \mathcal{T}_\mu \cup \bigcup_{i,j<\mu} \mathcal{T}_{ij}$. Let $T_\mu = \bigcap \mathcal{T}_\mu$, and for $i, j < \mu$ let $T_{ij} = \bigcap \mathcal{T}_{ij}$. It follows that $T = \bigcap \mathcal{T} = T_\mu \cap \bigcap_{i,j<\mu} T_{ij}$. The last term is an intersection of at most $|\mu^2 + 1| < \kappa$ theories. Moreover, $T \neq T_\mu$ and $T \neq T_{ij}$, for all $i, j < \mu$, since otherwise $T \cap A \neq \varnothing$. Thus, $T$ cannot be $\kappa$-prime. It follows that every $\kappa$-prime theory is a complete theory. $\square$

It is clear that the defining condition of $\omega$-disjunction is equivalent with the condition of disjunction in Definition 2.3. The proof of Theorem 2.7 is, in the special case $\kappa = \omega$, a new proof of the above mentioned well-known fact that in distributive lattices the meet-irreducible subsets are precisely the prime filters (see [23]).[11]

---

[8]That is, the condition $\bigvee_\kappa A \in T \Leftrightarrow A \cap T \neq \varnothing$, for any set $A$ of expressions such that $|A| < \kappa$, holds exactly for all $\kappa$-prime theories $T$ (and not only for the totally prime theories).

[9]This crucial property was not guaranteed in the proof of Theorem 3.4 [20] where we defined sets $\mathcal{T}_i = \{T' \in \mathcal{T} \mid a_i \notin T'\}$ satisfying a weaker condition. This problem in the proof of Theorem 3.4 [20] is hereby corrected.

[10]Note that some of these sets may be empty. The intersection of such an empty set is, by definition, the set of all formulas $Expr_\mathcal{L}$.

Suppose that $\mathcal{L}$ has also connectives for conjunction, intuitionistic implication and intuitionistic negation with the respective truth conditions given in Definition 2.3. It is not hard to check that the whole set of theories $Th_\mathcal{L}$ is stable with respect to the connectives of conjunction, intuitionistic implication and intuitionistic negation. This is shown in detail in the proof of Theorem 3.4 [20]. Consequently, Theorem 2.7 implies that the set of complete theories is exactly the set of all ($\omega$-) prime theories in the intuitionistic abstract logic $\mathcal{L}$.

The most prominent example of an intuitionistic abstract logic is the usual intuitionistic propositional logic developed by Brouwer, Heyting and Kolmogorov. Let us refer to this logic as BHK. BHK has a finitary relation of derivation $\vdash$ which, viewed as a compact closure operator, gives rise to a corresponding non-topped intersection structure. Because of compactness of $\vdash$ (see Fact 2.4), this intersection structure is minimally generated and the set of totally prime theories is the minimal generator set. We know that the set of complete theories is precisely the set of prime theories: $PTh_{BHK} = CTh_{BHK}$. Using the fact that intuitionistic Kripke semantics is sound and complete w.r.t. BHK, one checks that the conditions (ii) of Definition 2.3 are satisfied. Thus, BHK is in fact an intuitionistic abstract logic. Recall that BHK has the disjunction property, i.e. $\vdash a \curlyvee b$ implies $\vdash a$ or $\vdash b$. Thus, the smallest theory (this is the intersection of all theories, i.e. the set of all tautologies) is a prime theory $T_0$. Of course, this theory, which is generated by the set of all theories distinct from $T_0$, cannot be totally prime. By Zorn's Lemma, every theory extends to a maximal theory (a maximally consistent set). A maximal theory $T$ is the set of formulas forced by a world at a maximal position of a Kripke model. By the truth conditions of a Kripke model, $a \in T$ or $\sim a \in T$. Thus, every formula of the form $\sim a \curlyvee a$ is an element of every maximal theory and is therefore in the intersection of all maximal theories. This intersection cannot be the smallest theory, i.e. the set of all valid formulas, since $\sim a \curlyvee a$ is not valid in BHK. Thus, there are totally prime theories which are not maximal. This shows that the sets of maximal, totally prime and prime theories are pairwise distinct in BHK. In a classical abstract logic $\mathcal{L}$, however, these three sets collapse: $MTh_\mathcal{L} =$

---

[11]Let $(L, \curlywedge, \curlyvee)$ be a distributive lattice. Then $\mathcal{L} = (L, Th_\mathcal{L}, \{\curlyvee\})$ is a minimally generated abstract logic with disjunction, where $Th_\mathcal{L}$ is the set of proper filters on $L$. The logic is minimally generated because the union of any non-empty chain of filters (theories) is again a filter (a theory); see Fact 2.4 above. Theorem 2.7, with $\kappa = \omega$, now says that the set of complete theories (those theories $T$ with the property $a \curlyvee b \in T$ iff $a \in T$ or $b \in T$, for all $a, b$), i.e. the set of all prime filters, is exactly the set of all $\omega$-prime theories, i.e. the set of all meet-irreducible subsets of $L$.

$TPTh_\mathcal{L} = PTh_\mathcal{L}$. Indeed, a prime theory $T$ is generated in $\mathcal{L}$ by a set of maximal theories (recall that the set of maximal theories is the minimal generator set – by Definition 2.3 of a classical abstract logic), consequently $\sim a \curlyvee a \in T$ for every formula $a$. Thus, $\sim a \in T$ or $a \in T$, for any formula $a$. It follows that $T$ is not contained in any other theory, i.e. it is a maximal theory.

## 3  $B_4$, $K_3$ and $P_3$ abstract logics

In this section we define abstract logics that capture the essential semantic features of 4-valued logic $B_4$ and the 3-valued logics $K_3$ and $P_3$, respectively. That is, the well-known propositional logics of Dun/Belnap, Kleene and Priest will be particular cases of these abstract logics. In order to model these logics on an abstract level we go out from their respective truth-tables. Having only these truth-table informations, we cannot *a priori* assume that the corresponding abstract logics have compact consequence relations or are minimally generated. In particular, we cannot define these abstract logics by means of the minimal generator sets such as done in the intuitionistic case in Definition 2.3. Instead, we use the truth-tables in order to define certain *complete theories* which are contained in an ambient classical abstract logic. Then we will be able to prove that our abstract logics of type $B_4$ have in fact compact equivalence relations and can be represented as minimally generated abstract logics similarly as in the intuitionistic case (see Definition 2.3 above). Abstract logics that capture 4-valued Dunn/Belnap logic are studied by Font [7] in the context of the research field of Abstract Algebraic Logic. Font defines and characterizes these logics as *full models* of the class of De Morgan lattices. This essentially means that De Morgan lattices can be seen as the algebraic counterpart of Dunn/Belnap logic. We are able to show that our abstract logics of type $B_4$ coincide precisely with the class of *full models* studied in [7].

We introduce here $B_4$, $K_3$ and $P_3$ abstract logics in a similar way as the three-valued abstract logics presented in Definition 5.10 of [16] using only the corresponding truth-table informations:

**Definition 3.1** *Let $\mathcal{L} = (Expr_\mathcal{L}, Th_\mathcal{L}, \{\curlywedge, \curlyvee, \sim, \rightarrowtail\})$ be a classical abstract logic. Three further abstract logics $B_4(\mathcal{L})$, $K_3(\mathcal{L})$ and $P_3(\mathcal{L})$ are defined in terms of pairs $(A, \overline{A})$ of sets of $\mathcal{L}$-expressions. Let $A, \overline{A} \subseteq Expr_\mathcal{L}$ such that for all $a, b \in Expr_\mathcal{L}$ the following hold:*

*(i) $a \curlyvee b \in A \Leftrightarrow a \in A$ or $b \in A$; $a \curlyvee b \in \overline{A} \Leftrightarrow a \in \overline{A}$ and $b \in \overline{A}$*

*(ii) $a \curlywedge b \in A \Leftrightarrow a \in A$ and $b \in A$; $a \curlywedge b \in \overline{A} \Leftrightarrow a \in \overline{A}$ or $b \in \overline{A}$*

*(iii)* $\sim a \in A \Leftrightarrow a \in \overline{A}$; $\sim a \in \overline{A} \Leftrightarrow a \in A$

*(iv)* $a \rightarrowtail b \in A \Leftrightarrow a \in \overline{A}$ or $b \in A$; $a \rightarrowtail b \in \overline{A} \Leftrightarrow a \in A$ and $b \in \overline{A}$.

*Then the set $A$ is a complete $B_4$-theory (relative to $\mathcal{L}$) and $\overline{A}$ is its $B_4$-complement. If $A$ is $\mathcal{L}$-consistent (i.e., is contained in some $\mathcal{L}$-theory), then $A$ is a complete $K_3$-theory (relative to $\mathcal{L}$) and $\overline{A}$ is the $K_3$-complement of $A$. If $A$ contains a complete (i.e. a maximal) $\mathcal{L}$-theory, then $A$ is a complete $P_3$-theory (relative to $\mathcal{L}$) and $\overline{A}$ is its $P_3$-complement. The abstract logic generated by the set of all complete $B_4$-theories is $B_4(\mathcal{L}) = (Expr_\mathcal{L}, Th_{B_4(\mathcal{L})}, \{\curlywedge, \curlyvee, \sim, \rightarrowtail\})$, where $Th_{B_4(\mathcal{L})}$ is the set of all intersections of non-empty sets of complete $B_4$-theories. Similarly, we define the abstract logics $K_3(\mathcal{L})$ and $P_3(\mathcal{L})$. The class of all $B_4$ abstract logics is given by $\{B_4(\mathcal{L}) \mid \mathcal{L}$ is any classical abstract logic$\}$. Similarly, we define the classes of $K_3$ abstract logics and of $P_3$ abstract logics. We refer to these logics also as abstract logics of type $B_4$ (of type $K_3$, of type $P_3$), respectively.*

Intuitively, a complete theory $A$ assigns truth values to formulas: elements of $A$ are the formulas with truth value *true* while the elements of its complement $\overline{A}$ are the formulas with truth value *false*. Note that the above definition derives from the truth-table informations of the respective many-valued propositional logics of Dunn/Belnap, Kleene and Priest.

Let $A$ be a complete $B_4$-, $K_3$-, or $P_3$-theory. Observe that
$a \rightarrowtail b \in A$ iff $\sim a \curlyvee b \in A$,
$\sim (a \curlywedge b) \in A$ iff $\sim a \curlyvee \sim b \in A$,
$\sim (a \curlyvee b) \in A$ iff $\sim a \curlywedge \sim b \in A$,
$a \in A$ iff $\sim\sim a \in A$,
$a \curlywedge b \in A$ iff $\sim (\sim a \curlyvee \sim b) \in A$,
$a \curlyvee b \in A$ iff $\sim (\sim a \curlywedge \sim b) \in A$. That is, the respective formulas are logically equivalent in these abstract logics. This shows in particular that we can work without the connective of implication.

Let $CTh_{B_4(\mathcal{L})}$, $CTh_{K_3(\mathcal{L})}$, $CTh_{P_3(\mathcal{L})}$ denote the set of complete theories of logic $B_4(\mathcal{L})$, $K_3(\mathcal{L})$, $P_3(\mathcal{L})$, respectively. Then follows that

- $CTh_{K_3(\mathcal{L})} \cup CTh_{P_3(\mathcal{L})} \subseteq CTh_{B_4(\mathcal{L})}$,

- $CTh_{K_3(\mathcal{L})} \cap CTh_{P_3(\mathcal{L})} = MTh_\mathcal{L} = CTh_\mathcal{L}$.

Given a complete theory $A$ of any abstract logic, we interpret the elements of $A$ as the formulas which are *true*. We define *falsity* as *negation*, whenever an adequate negation is given. Our notion of *adequacy of negation* is given in the following definition.

**Definition 3.2** *Let $\mathcal{L}$ be any (classical or non-classical) abstract logic with an unary connective $\sim$ and the set $CTh_\mathcal{L}$ of complete theories (i.e. the greatest set of theories stable w.r.t. the connectives). We call $\sim$ an adequate negation of $\mathcal{L}$ if $b \in T$ implies $\sim\sim b \in T$, for all $b \in Expr_\mathcal{L}$ and for all $T \in CTh_\mathcal{L}$. If $\sim$ is an adequate negation, then for $b \in Expr_\mathcal{L}$ and $T \in CTh_\mathcal{L}$ we say that*

- *$b$ is true (w.r.t. $T$) if $b \in T$,*
- *$b$ is false (w.r.t. $T$) if $\sim b \in T$.*

An adequate negation $\sim$ satisfies the following basic intuition: "If $b$ is true, then the negation of $b$ is false. If $b$ is false, then the negation of $b$ is true." Note that classical and all non-classical negations considered so far are adequate.

The proof of the next result is an easy exercise.

**Lemma 3.3** *Let $\mathcal{L}$ be a classical abstract logic and $A \subseteq Expr_\mathcal{L}$. Then $A$ is a complete $B_4(\mathcal{L})$-theory iff the following conditions hold for all $a, b \in Expr_\mathcal{L}$:*

*(i) $a \in A \Leftrightarrow \sim\sim a \in A$,*

*(ii) $a \curlyvee b \in A \Leftrightarrow a \in A$ or $b \in A$,*

*(iii) $\sim(a \curlyvee b) \in A \Leftrightarrow \sim a \in A$ and $\sim b \in A$,*

*(iv) $a \curlywedge b \in A \Leftrightarrow a \in A$ and $b \in A$,*

*(v) $\sim(a \curlywedge b) \in A \Leftrightarrow \sim a \in A$ or $\sim b \in A$,*

*Similarly, $A$ is a complete $K_3(\mathcal{L})$-theory iff $A$ is $\mathcal{L}$-consistent and satisfies the above conditions (i) – (v); and $A$ is a complete $P_3(\mathcal{L})$-theory iff $A$ contains a complete $\mathcal{L}$-theory, and the conditions (i) – (v) are satisfied.*

Taking into account the last result one recognizes that our abstract logics of type $B_4$ are very closely related to *De Morgan Logics* such as studied by Beziau [1] (note that we abstract from any syntactical structure). We will use semantic methods in order to show compactness of the consequence relation (Theorem 3.6 below). Note that factorizing formulas modulo logically equivalence one obtains in fact a De Morgan lattice.

For the following result we will need for the first time the compactness of the ambient classical abstract logic $\mathcal{L}$.[12]

---

[12]By compactness of $\mathcal{L}$ we mean here the compactness of its consequence relation. There are further notions of compactness in a logic (see [12], [20]) which in the case of classical abstract logics are equivalent to compactness of the consequence relation.

**Theorem 3.4** *Let $\mathcal{L}$ be a classical abstract logic and let $A \subseteq Expr_\mathcal{L}$ be a complete $B_4(\mathcal{L})$-theory.*

(i) *$A$ is a complete $K_3(\mathcal{L})$-theory iff $A$ contains no $\mathcal{L}$-contradiction (i.e. no formula which is inconsistent in logic $\mathcal{L}$).*

(ii) *$A$ is a complete $P_3(\mathcal{L})$-theory iff $A$ contains all $\mathcal{L}$-tautologies (i.e. all formulas which are contained in all theories of logic $\mathcal{L}$).*

**Proof.** (i): The direction from left to right is trivial. Suppose that the complete $B_4$-theory $A$ contains no $\mathcal{L}$-contradiction. We must show that $A$ is $\mathcal{L}$-consistent. Suppose not. By compactness of $\mathcal{L}$, there is a finite $\mathcal{L}$-inconsistent subset $A_f \subseteq A$. Let $A_f = \{a_0, ..., a_n\}$. Then the formula $b = (...(a_0 \curlywedge a_1) \curlywedge a_2) \curlywedge ... \curlywedge a_n)$ is an $\mathcal{L}$-contradiction. Since $A$ is stable under conjunction, one easily checks that $b \in A$. This is a contradiction. Thus, $A$ must be $\mathcal{L}$-consistent.

(ii): Again, the direction from left to right is trivial. Suppose that the complete $B_4$-theory $A$ contains all $\mathcal{L}$-tautologies. We must show that $A$ contains a complete (i.e. a maximal) $\mathcal{L}$-theory. Suppose not. Then for each complete $\mathcal{L}$-theory $T$ we may choose a formula $a_T \in T \smallsetminus A$. The set $\{\sim a_T \mid T \in MTh_\mathcal{L}\}$ is $\mathcal{L}$-inconsistent and has, by compactness, a finite $\mathcal{L}$-inconsistent subset $B = \{\sim a_{T_0}, ..., \sim a_{T_n}\}$. Then the formula $(...(\sim a_{T_0} \curlywedge \sim a_{T_1}) \curlywedge \sim a_{T_2}) \curlywedge ... \curlywedge \sim a_{T_n})$ is an $\mathcal{L}$-contradiction. Consequently, the formula $(...(a_{T_0} \curlyvee a_{T_1}) \curlyvee a_{T_2}) \curlyvee ... \curlyvee a_{T_n})$ is an $\mathcal{L}$-tautology and belongs to $A$. Since $A$ is stable under disjunction, it follows that $a_{T_0} \in A$ or $a_{T_1} \in A$ or ... or $a_{T_n} \in A$. This contradiction shows that $A$ must contain some complete $\mathcal{L}$-theory. $\square$

In [16] we defined the complete $P_3$-theories in the same way as in item (ii) of Theorem 3.4 whereas the complete $K_3$-theories were defined as in Definition 3.1 of the present paper. This was somewhat unsatisfactory since it occults the dual character of both logics. This duality is now clear by Definition 3.1 and by the characterizations of $P_3$ and $K_3$ given in Theorem 3.4.

The next result was already proved in [16] (for the logics $K_3$ and $P_3$). It is interesting that we do not need here compactness of the consequence relations in order to show existence of the respective minimal generator sets (see Fact 2.4).

**Corollary 3.5** *Let $\mathcal{L}$ be a classical abstract logic. Then the abstract logics $B_4(\mathcal{L})$, $K_3(\mathcal{L})$ and $P_3(\mathcal{L})$ are minimally generated.*

**Proof.** The set of all complete $B_4(\mathcal{L})$-theories is a generator set of $B_4(\mathcal{L})$. From Lemma 3.3 it follows that the set of complete $B_4(\mathcal{L})$-theories is closed under union of non-empty chains. That is, if $\alpha > 0$ is any ordinal and $(A_i \mid i < \alpha)$ is a chain of complete $B_4(\mathcal{L})$-theories, then $\bigcup_{i<\alpha} A_i$ is a complete $B_4(\mathcal{L})$-theory. By Fact 2.5, this chain condition is sufficient for the existence of a minimal generator set. Using Theorem 3.4, one recognizes that this chain condition also holds in $K_3(\mathcal{L})$ and $P_3(\mathcal{L})$. $\square$

**Theorem 3.6** *Let $\mathcal{L}$ be a classical abstract logic. The consequence relation of the abstract logic $B_4(\mathcal{L})$ is compact.*

**Proof.** Let $cl(.)$ be the closure operator on $Expr_\mathcal{L}$ associated with the consequence relation of $B_4(\mathcal{L})$. It is a well-known fact (see, e.g., Theorem 7.14 [5]) that $cl$ is algebraic (i.e., compact) iff for every directed family $\{A_i\}_{i\in I}$ of subsets of $Expr_\mathcal{L}$ it holds that $cl(\bigcup_{i\in I} A_i) = \bigcup_{i\in I} cl(A_i)$. The complete theories form a generator set. Thus, for any set of expressions $A$ and any expression $a$ it holds the following: $a \in cl(A)$ iff every complete $B_4(\mathcal{L})$-theory that contains $A$ also contains $a$. We can close $A$ under the connectives. Let $A'$ denote this closure. That is, $A'$ is the smallest set containing $A$ and satisfying $b \in A'$ iff $\sim\sim b \in A'$, $b \curlyvee c \in A'$ iff ($b \in A'$ or $c \in A'$), ... etc. for any $b, c$. Then $A'$ is the smallest complete $B_4(\mathcal{L})$-theory containing $A$. It follows that $a \in cl(A)$ iff $a \in A'$, that is, $cl(A) = A'$. Thus, it is enough to show that $(\bigcup_{i\in I} A_i)' = \bigcup_{i\in I} A'_i$. Of course, for each $j \in I$, $A'_j \subseteq (\bigcup_{i\in I} A_i)'$. Hence, $\bigcup_{i\in I} A'_i \subseteq (\bigcup_{i\in I} A_i)'$. From Lemma 3.3 it follows that the union of complete $B_4(\mathcal{L})$-theories is again a complete $B_4(\mathcal{L})$-theory. Thus, $\bigcup_{i\in I} A'_i$ is a complete $B_4(\mathcal{L})$-theory containing all $A_i$. But $(\bigcup_{i\in I} A_i)'$ is the smallest complete $B_4(\mathcal{L})$-theory containing all $A_i$. It follows that $\bigcup_{i\in I} A'_i = (\bigcup_{i\in I} A_i)'$. $\square$

We are now able to characterize $B_4$ abstract logics by means of the minimal generator set and compactness of its consequence relation, independently from the ambient classical abstract logic.

**Theorem 3.7** *Let $\mathcal{L} = (Expr_\mathcal{L}, Th_\mathcal{L}, \{\curlyvee, \curlywedge, \sim\})$ be an abstract logic. Then $\mathcal{L}$ is a $B_4$ abstract logic iff $\mathcal{L}$ has a compact consequence relation,[13] $\{\varnothing, Expr_\mathcal{L}\} \subseteq Th_\mathcal{L}$, and for all $A \in TPTh_\mathcal{L}$ and all $a, b \in Expr_\mathcal{L}$ the following truth conditions are satisfied:*

(i) $a \in A \Leftrightarrow \sim\sim a \in A$,

(ii) $a \curlyvee b \in A \Leftrightarrow a \in A$ *or* $b \in A$,

(iii) $\sim(a \curlyvee b) \in A \Leftrightarrow \sim a \in A$ *and* $\sim b \in A$,

---
[13] In particular, $\mathcal{L}$ is minimally generated and $TPTh_\mathcal{L}$ is the smallest generator set; see Fact 2.4.

*(iv)* $a \curlywedge b \in A \Leftrightarrow a \in A$ *and* $b \in A$,

*(v)* $\sim (a \curlywedge b) \in A \Leftrightarrow \sim a \in A$ *or* $\sim b \in A$.

**Proof.** Let $\mathcal{L}$ be a $B_4$ abstract logic. By Theorem 3.6, the consequence relation is compact. The smallest generator set $TPTh_\mathcal{L}$ is contained in every generator set, in particular in the set of all complete $B_4$-theories. Hence, by Lemma 3.3, (i) – (v) are satisfied. In particular, $\varnothing$ and $Expr_\mathcal{L}$ are complete theories. Now suppose that $\mathcal{L}$ is an abstract logic with compact consequence relation, $\{\varnothing, Expr_\mathcal{L}\} \subseteq Th_\mathcal{L}$ and truth conditions (i) – (v). Then the set of all theories satisfying (i) – (v) is the set of complete theories. Note that in particular the empty set and $Expr_\mathcal{L}$ are complete theories. The ambient classical abstract logic $\mathcal{L}'$ is obtained by choosing as generator set the collection of those complete theories $A$ having the additional property: $a \curlyvee \sim a \in A$ and $a \curlywedge \sim a \notin A$, for all expressions $a$. Let us show that the induced consequence relation $\Vdash_{\mathcal{L}'}$ is compact: Suppose $A \Vdash_{\mathcal{L}'} a$. We close $A$ under the connectives in $\mathcal{L}$ according to the conditions (i) – (v) and get its closure $A'$ (see the proof of Theorem 3.6). Obviously, this is the smallest complete $\mathcal{L}$-theory containing $A$. By compactness of the consequence relation $\Vdash_\mathcal{L}$ of $\mathcal{L}$, there is a finite $A_f \subseteq A$ such that $A_f \Vdash_\mathcal{L} a$. Since every complete theory of logic $\mathcal{L}'$ is a complete theory of logic $\mathcal{L}$, it follows that $A_f \Vdash_{\mathcal{L}'} a$. Thus, $\Vdash_{\mathcal{L}'}$ is compact. We have restored the classical abstract logic $\mathcal{L}'$ such that $\mathcal{L} = B_4(\mathcal{L}')$ according to Definition 3.1. $\square$

**Corollary 3.8** *Let $\mathcal{L}$ be a $B_4$ abstract logic. Then the set of complete theories is precisely the set of all prime theories: $CTh_\mathcal{L} = PTh_\mathcal{L}$.*

**Proof.** By Theorem 2.7, all complete theories are $\omega$-prime, since they satisfy truth condition (ii) of Theorem 3.7. Thus, $CTh_\mathcal{L} \subseteq PTh_\mathcal{L}$. Now let $A$ be a $\omega$-prime theory. $A$ is the intersection of a non-empty set of totally prime theories. We apply the truth condition satisfied by the totally prime theories: $a \in A$ iff $a$ is contained in every totally prime theory extending $A$ iff $\sim\sim a$ is contained in every totally prime theory extending $A$ iff $\sim\sim a \in A$. Similarly one shows condition (iii) for all prime $A$. Condition (ii) holds by Theorem 2.7. Condition (iv) holds for all theories. Finally, for any prime $A$: $\sim (a \curlywedge b) \in A$ iff $\sim (a \curlywedge b)$ is contained in all totally prime theories extending $A$ iff every totally prime theory extending $A$ contains $\sim a$ or contains $\sim b$ iff every totally prime theory extending $A$ contains $\sim a \curlyvee \sim b$ iff $(\sim a \curlyvee \sim b) \in A$. Thus, all prime theories are complete.[14] $\square$

---

[14]Note that our argument shows that not only prime theories but all theories satisfy the truth conditions (i), (iii), (iv) and (v), whereas (ii) is only satisfied by prime theories. That is, if we had no disjunction, then the set of complete theories would be the set of all theories.

Now we may establish a connection to the approach of Font [7], where a certain class of abstract logics is defined as the class of *full models* of the class of De Morgan lattices **DM**. Font uses this class of abstract logics in order to show that **DM** is, in some precise sense, the algebraic counterpart of Dunn/Belnap logic $B_4$. The definition of *full model of **DM*** presented in [7] is not easily accessible for a reader not familiar with the sophisticated underlying algebraic theory. However, it turns out that our class of $B_4$ abstract logics, whose definition is based on simpler algebraic assumptions, coincides with the class of full models of **DM** as defined in [7].

**Theorem 3.9** *An abstract logic $\mathcal{L}$ is a $B_4$ abstract logic iff $\mathcal{L}$ is a full model of the class of De Morgan lattices in the sense of [7].*

**Proof.** We consider the characterization of *full model* given in Theorem 4.3 [7]. If $\mathcal{L}$ is a $B_4$ abstract logic, then, by Theorem 3.6, the consequence relation is compact (in the language of [7]: the logic is finitary) and the set of all complete theories is, by definition, a generator set. By Corollary 3.8, this is the set of all prime theories. Note that in particular the empty set $\varnothing$ as well as the whole set of expressions $Expr_\mathcal{L}$ are prime theories. As pointed out on p. 16 in [7], the function $\Phi(.)$ on $Pow(Expr_\mathcal{L})$, defined by $\Phi(A) = Expr_\mathcal{L} \smallsetminus \{\sim a \mid a \in A\}$ and used in Theorem 4.3 [7], maps prime filters to prime filters and satisfies for any prime filter $T$ the equation $\Phi(\Phi(T)) = T$.[15] Thus, the conditions of Theorem 4.3 [7] are satisfied and $\mathcal{L}$ is a full model of the class of De Morgan lattices in the sense of [7].
The other way arround, suppose that an abstract logic $\mathcal{L}$ satisfies the conditions given in Theorem 4.3 [7]. Then $\varnothing$ is a closure, i.e. a theory in our sense. It follows that $Expr_\mathcal{L}$ is also a theory, i.e. there are no contradictory formulas (otherwise, the corresponding De Morgan lattice would be bounded and would have a greatest element and $\varnothing$ would not be a closure). Furthermore, the consequence relation is compact. By Fact 2.4, the logic is minimally generated. Since $TPTh_\mathcal{L}$ is the smallest generator set, the set $TPTh_\mathcal{L} \smallsetminus \{Expr_\mathcal{L}\}$ is contained in any basis of the abstract logic. Thus, the elements of $TPTh_\mathcal{L} \smallsetminus \{Expr_\mathcal{L}, \varnothing\}$ satisfy the conditions (1), (2) and (3) of Theorem 4.3 [7]. In view of Theorem 3.7, it is enough to show that the elements of $TPTh_\mathcal{L}$ satisfy the truth conditions (i) – (v). The conditions (ii) and (iv) correspond to the conditions (2) and (1) of Theorem 4.3 [7], respectively. In order to show (i), (iii) and (v), we use the properties of the function $\Phi$ given in condition (3) of Theorem 4.3 [7]. For any expression $a$ and any prime filter $A \subseteq Expr_{\mathcal{L}}$: $a \in A = \Phi(\Phi(A)) \Leftrightarrow\, \sim a \notin \Phi(A) \Leftrightarrow\, \sim\sim a \in A$. Thus, (i) holds. Similarly, $\sim (a \curlywedge b) \in A \Leftrightarrow\, \sim\sim (a \wedge b) \notin \Phi(A) \Leftrightarrow a \curlywedge b \notin \Phi(A) \Leftrightarrow (a \notin \Phi(A)$ or $b \notin \Phi(A)) \Leftrightarrow (\sim a \in A$ or $\sim b \in A) \Leftrightarrow\, \sim a \curlyvee \sim b \in A$, for all expressions $a, b$ and all prime filters $A \subseteq Expr_{\mathcal{L}}$. Thus, condition (iii) is satisfied. Condition (v) of Theorem 3.7 follows similarly. □

---

[15]Note that the set of prime filters is given by $PTh(\mathcal{L}) \smallsetminus \{\varnothing, Expr_\mathcal{L}\}$. Also recall that the notion of basis of an abstract logic may differ from our notion of generator set if the whole set of formulas is consistent, i.e. a theory. In this case, a generator set necessarily contains the whole set of formulas while a basis generates this theory by intersection of the empty set of theories.

We finish this study on abstract logics with an exercise showing that the well-known propositional logics $B_4$, $K_3$ and $P_3$ are in fact particular $B_4$, $K_3$ and $P_3$ abstract logics, respectively. That is, we show that these propositional logics have the corresponding features given in Definition 3.1. For this we suppose that $\mathcal{L}$ is the classical propositional logic. That is, the expressions $Expr_{\mathcal{L}}$ are the usual propositional formulas (with connectives $\{\curlywedge, \curlyvee, \sim, \rightarrowtail\}$) inductively defined over a countable infinite set $P$ of propositional variables, and the generator set is the set of maximal theories. These maximal theories can be given in a proof-theoretic way as the maximally consistent sets of an underlying deductive system or in a semantic way as sets satisfied by valuations $v : P \to \{0, 1\}$ (of course, the valuations extend to the set of all formulas). We consider here the semantic approach. Classical propositional logic is compact and so it is clear that $\mathcal{L}$ can be seen as a classical abstract logic in our sense. Let us look at the abstract logic $K_3(\mathcal{L})$ and the usual propositional logic $K_3$. In order to show the desired correspondence it is sufficient to establish a bijection $A \mapsto v$ between the complete $K_3(\mathcal{L})$-theories $A$, defined in our sense, and the usual $K_3$-valuations $v : P \to \{0, 1, i\}$ such that $A$ and $v$ assign the same truth value to every formula $b \in Expr_{\mathcal{L}}$. $\{0, 1, i\}$ is the set of truth values in $K_4$, and $\{1\}$ is the set of designated values. $v$ extends to the set of all formulas according to the $K_3$ truth tables. Let $A = \{b \in Expr_{\mathcal{L}} \mid v(b) = 1\}$ and $\overline{A} = \{b \in Expr_{\mathcal{L}} \mid v(b) = 0\}$. One easily checks that $A$ is consistent in logic $\mathcal{L}$, and $A$ and $\overline{A}$ satisfy the conditions of Definition 3.1. Thus, $A$ is a complete $K_3(\mathcal{L})$-theory. Conversely, suppose $A$ is a complete $K_3(\mathcal{L})$-theory. We define a valuation $v : P \to \{0, 1, i\}$ by

$$v(p) = \begin{cases} 1, & \text{if } p \in A \\ 0, & \text{if } \sim p \in A \\ i, & \text{else} \end{cases}$$

Now we extend $v$ to the set of all formulas according to the $K_3(\mathcal{L})$ truth tables. Then one shows inductively: $v(b) \in \{1\} \Leftrightarrow b \in A$, and $v(b) = 0 \Leftrightarrow\, \sim b \in A$, for all formulas $b$. This yields the desired one-to-one correspondence between the valuations of propositional logic $K_3$ and the complete $K_3(\mathcal{L})$-theories defined in our sense. Similarly, one shows that the propositional logic $P_3$ (see, for instance, [22]) is a $P_3$ abstract logic. Recall that in this case the designated set is $\{1, i\}$. In order to establish a bijection $A \mapsto v$ between the complete $P_3(\mathcal{L})$-theories of $\mathcal{L}$ and the valuations $v : P \to \{1, 0, i\}$ one defines for a given valuation $v$ the sets $A = \{b \mid v(b) \in \{1, i\}\}$ and $\overline{A} = \{b \mid v(b) \in \{i, 0\}\}$. Then $A$ is a complete $P_3(\mathcal{L})$-theory with complement $\overline{A}$, and $A$ and $v$ give rise to the same truth values. Conversely, for a given complete $P_3(\mathcal{L})$-theory $A$ one defines a $P_3$-valuation $v$ by

$$v(p) = \begin{cases} 1, & \text{if } p \in A \text{ and } \sim p \notin A \\ 0, & \text{if } \sim p \in A \text{ and } p \notin A \\ i, & \text{if } p \in A \text{ and } \sim p \in A \end{cases}$$

Notice that the case $a \notin A$ and $\sim a \notin A$ is impossible, since $a \curlyvee \sim a$ is valid in classical propositional logic $\mathcal{L}$ and is therefore an element of $A$. But this implies $a \in A$ or $\sim a \in A$. Extending $v$ to the set of all formulas one gets the following for all formulas $b$: $v(b) = 1 \Leftrightarrow (b \in A$ and $\sim b \notin A)$; $v(b) = i \Leftrightarrow (b \in A$ and $\sim b \in A)$; $v(b) = 0 \Leftrightarrow (\sim b \in A$ and $b \notin A)$. That is, $v$ and $A$ assign precisely the same truth values to $b$. Similarly, one shows the desired correspondence between the complete $B_4(\mathcal{L})$-theories and the valuations of 4-valued propositional logic $B_4$. Note that a valuation is here a function $v : P \to \{0, 1, N, B\}$, where the designated values are $\{1, B\}$.

## 4 Syntax of the $\in_T$-style extensions

Let $\mathcal{L} = (Expr_\mathcal{L}, Th_\mathcal{L}, \{\curlywedge, \curlyvee, \sim, \rightarrowtail\})$ be a classical abstract logic. We refer to the elements of $Expr_\mathcal{L}$ as the $\mathcal{L}$-expressions or $\mathcal{L}$-formulas. The alphabet of the $\in_T$-extension is given by the $\mathcal{L}$-expressions, a (possibly empty) set $C$ of constant symbols (distinct from the symbols of $\mathcal{L}$), a countable infinite set of variables $V = \{v_0, v_1, v_2, ...\}$ which is well-ordered by the given enumeration, logical connectives $\vee, \wedge$ and $: false$, predicates (operator symbols) for truth and falsity : $true$, $: false$, respectively (we use postfix notation), the identity connective $\equiv$, the reference connective $<$, quantifiers $\exists$ and $\forall$ and auxiliary symbols: $)$, $($, and dot. The operator $: false$ for the falsity predicate is also viewed as the logical connective for (classical or non-classical) negation. Recall that we interpret negation as falsity (see Definition 3.2). The sets $C$ and $Expr_\mathcal{L}$ are viewed as parameter sets. We refer to the logics $\mathcal{L}$, $B_4(\mathcal{L})$, $K_3(\mathcal{L})$ and $P_3(\mathcal{L})$ as parameter logics. We will see that each of these parameter logics can be extended by a corresponding $\in_T$-style logic. This parametrized logic turns out to be again a classical, $B_4$, $K_3$ or $P_3$ abstract logic.

**Definition 4.1** *The set of expressions (or formulas) $Expr(\mathcal{L}, C)$ is the smallest set that contains $V \cup C$ and is closed under the following condition. If $\varphi, \psi \in Expr(\mathcal{L}, C)$, then $(\varphi : true)$, $(\varphi : false)$, $(\varphi \vee \psi)$, $(\varphi \wedge \psi)$, $(\varphi \equiv \psi)$, $(\varphi < \psi)$, $\exists x.\varphi$, $\forall x.\varphi \in Expr(\mathcal{L}, C)$.*

The notions of subformula and the set of free variables of a formula are defined in the obvious way. By $sub(\varphi)$, $fvar(\varphi)$, $var(\varphi)$, $con(\varphi)$, $fc\mathcal{L}(\varphi)$, $\mathcal{L}(\varphi)$ we denote the set of subformulas of $\varphi$, the set of free variables, of variables, of constant symbols, of $\mathcal{L}$-expressions occurring in $\varphi$, respectively. Furthermore, $fc\mathcal{L}(\varphi) := fvar(\varphi) \cup con(\varphi) \cup \mathcal{L}(\varphi)$. Note that according to our definition strings such as $\exists x.c$ or $\forall x.\exists x.x$ are formulas. This is somewhat unintuitive in our intensional setting where the denotation of a formula may correspond to the intension, i.e. the sense, that the formula expresses. For instance, the semantic truth conditions imply that $c < \exists x.c$ is valid. But $\exists x.c$ does not say anything meaningful about $c$. In order to avoid such counter-intuitive side effects we assume that $Expr(\mathcal{L}, C)$ contains only formulas $\varphi$ with the following intended property: Whenever $\exists x.\psi$ or $\forall x.\psi$ is a subformula of $\varphi$, then $x \in fvar(\psi)$. In fact, one can give an inductive definition of the set of all *intended* formulas (see [18]).

## 4.1 Substitutions

**Definition 4.2** *A substitution is a function $\sigma : V \cup C \cup Expr_\mathcal{L} \to Expr(\mathcal{L}, C)$. If $A \subseteq V \cup C \cup Expr_\mathcal{L}$ and $\sigma(u) = u$ for all $u \in (V \cup C \cup Expr_\mathcal{L}) \smallsetminus A$, then we write $\sigma : A \to Expr(\mathcal{L}, C)$. If $\sigma$ is a substitution, $u_0, ..., u_n \in V \cup C \cup Expr_\mathcal{L}$ and $\varphi_0, ..., \varphi_n \in Expr(\mathcal{L}, C)$, then the substitution $\sigma[u_0 := \varphi_0, ..., u_n := \varphi_n]$ is defined by:*

$$\sigma[u_0 := \varphi_0, ..., u_n := \varphi_n](v) = \begin{cases} \varphi_i & \text{if } v = u_i, \text{ for some } i \leq n \\ \sigma(v) & \text{else} \end{cases}$$

*The identity substitution $u \mapsto u$ is denoted by $\varepsilon$. Instead of $\varepsilon[u_0 := \varphi_0, ..., u_n := \varphi_n]$ we also write $[u_0 := \varphi_0, ..., u_n := \varphi_n]$. A substitution $\sigma$ extends in the following way to a function $[\sigma] : Expr(\mathcal{L}, C) \to Expr(\mathcal{L}, C)$ (we use postfix notation*

*for $[\sigma]$):*

$$
\begin{aligned}
&u[\sigma] := \sigma(u), \text{ for } u \in V \cup C \cup Expr_{\mathcal{L}} \\
&(\varphi : true)[\sigma] := \varphi[\sigma] : true \\
&(\varphi : false)[\sigma] := \varphi[\sigma] : false \\
&(\varphi \vee \psi)[\sigma] := \varphi[\sigma] \vee \psi[\sigma] \\
&(\varphi \wedge \psi)[\sigma] := \varphi[\sigma] \wedge \psi[\sigma] \\
&(\varphi \equiv \psi)[\sigma] := \varphi[\sigma] \equiv \psi[\sigma] \\
&(\varphi < \psi)[\sigma] := \varphi[\sigma] < \psi[\sigma] \\
&(\exists x.\varphi)[\sigma] := \exists y.\varphi[\sigma[x := y]] \\
&(\forall x.\varphi)[\sigma] := \forall y.\varphi[\sigma[x := y]],
\end{aligned}
$$

*where $y = min(V \smallsetminus \bigcup\{fvar(\sigma(u)) \mid u \in fcon(\exists x.\varphi)\})$. We say that $y$ is forced by $\sigma$ w.r.t. $\exists x.\varphi$. For two substitutions $\sigma$ and $\tau$ the composition is the substitution $\sigma \circ \tau$ defined by $(\sigma \circ \tau)(u) = \sigma(u)[\tau]$, for $u \in V \cup C \cup Expr_{\mathcal{L}}$.*

**Definition 4.3** *To expressions $\varphi$ and $\psi$ are said to be alpha-congruent or $\alpha$-congruent, notation: $\varphi =_\alpha \psi$ if they differ at most on their bound variables.*

**Definition 4.4** *For $\varphi, \psi \in Expr(\mathcal{L}, C)$ we define $\varphi \prec \psi :\Leftrightarrow$ there are $x \in V$ and $\psi' \in Expr(\mathcal{L}, C) \smallsetminus \{x\}$ such that $x \in fvar(\psi')$ and $\psi'[x := \varphi] =_\alpha \psi$. The relation $\prec$ is called syntactical reference.*

The syntactical reference $\varphi \prec \psi$ expresses that formula $\psi$ "says something about" or "refers to" formula $\psi$. Technically, $\varphi \prec \psi$ iff $\varphi$ is alpha-congruent to a proper subformula $\varphi'$ of $\psi$, and every free occurrence of a variable in $\varphi'$ remains free in $\psi$. Here come some examples: $x \prec (x : true)$, $(\varphi \vee \psi) \prec \chi \to (\varphi \vee \psi)$, $y \prec \forall x.(x < y)$, but $x \not\prec \forall x.(x < y)$.

The syntactical reference $\prec$ is a transitive relation on the set of formulas. If $\varphi \prec \psi$ and $\sigma$ is a substitution, then $\varphi[\sigma] \prec \psi[\sigma]$. Further useful properties of substitutions, alpha-congruence and the syntactical reference can be found in [18].

## 5 Semantics of the $\in_T$-extensions

We are given a classical abstract logic $\mathcal{L}$ and a set $C$ of constant symbols (not occurring in $\mathcal{L}$). We consider the complete lattice of truth values $L = \{0, 1, B, N\}$ with the ordering $\leq_L$ given by $0 \leq_L B \leq_L 1$ and $0 \leq_L N \leq_L 1$. By $sup(X)$, $inf(X)$ we denote the supremum, the infimum in $L$, respectively, of a set $X \subseteq L$.

**Definition 5.1** *Let $T$ be a complete $B_4(\mathcal{L})$-theory.[16] A model over $T$ is a structure $\mathcal{M} = (M, TRUE, FALSE, <^{\mathcal{M}}, \Gamma, T)$ given by:*

- *a non-empty set $M$ of propositions (the elements of $M$ are generally given as abstract entities without any inner structure),*

- *a set $TRUE \subseteq M$ of true propositions and a set $FALSE \subseteq M$ of false propositions such that*

    - *$TRUE \cap FALSE = \varnothing$ iff $T$ is a $K_3$-theory*
    - *$M = TRUE \cup FALSE$ iff $T$ is a $P_3$-theory*

- *a transitive relation $<^{\mathcal{M}} \subseteq M \times M$ for semantical reference,*

- *a semantic function, called Gamma-function, $\Gamma : Expr(\mathcal{L}, C) \times M^V \to M$ that maps an expression $\varphi$ to its denotation $\Gamma(\varphi, \gamma) \in M$. $\Gamma$ depends on assignments $\gamma : V \to M$ of propositions to variables. If $\gamma \in M^V$ is an assignment and $\sigma$ is a substitution, then $\gamma\sigma \in M^V$ denotes the assignment defined by $x \mapsto \Gamma(\sigma(x), \gamma)$. If $x \in V, m \in M$, then $\gamma_x^m$ is the assignment defined by*

$$\gamma_x^m(y) := \begin{cases} m, \text{ if } x = y \\ \gamma(y), \text{ else.} \end{cases}$$

*The Gamma-function satisfies the following structure conditions:*

(EP) *For all $x \in V$ and all assignments $\gamma \in M^V$, $\Gamma(x, \gamma) = \gamma(x)$. (Extension Property)*

(CP) *If $\varphi \in Expr(\mathcal{L}, C)$, $\gamma, \gamma' \in M^V$, and $\gamma(x) = \gamma'(x)$ for all $x \in fvar(\varphi)$, then $\Gamma(\varphi, \gamma) = \Gamma(\varphi, \gamma')$. (Coincidence Property)[17]*

(SP) *If $\varphi \in Expr(\mathcal{L}, C)$, $\gamma \in M^V$ and $\sigma : V \to Expr(\mathcal{L}, C)$ is a substitution, then $\Gamma(\varphi[\sigma], \gamma) = \Gamma(\varphi, \gamma\sigma)$. (Substitution Property)*

(RP) *If $\varphi \prec \psi$, then $\Gamma(\varphi, \gamma) <^{\mathcal{M}} \Gamma(\psi, \gamma)$, for all $\varphi, \psi \in Expr(C)$ and all assignments $\gamma$. (Reference Property)*

---

[16]Recall that there are four possibilities for $T$: $T \in CTh_{K_3(\mathcal{L})} \smallsetminus CTh_{P_3(\mathcal{L})}, T \in CTh_{P_3(\mathcal{L})} \smallsetminus CTh_{K_3(\mathcal{L})}, T \in CTh_{\mathcal{L}} = CTh_{K_3(\mathcal{L})} \cap CTh_{P_3(\mathcal{L})}, T \in CTh_{B_4(\mathcal{L})} \smallsetminus (CTh_{K_3(\mathcal{L})} \cup CTh_{P_3(\mathcal{L})})$.

[17]If $fvar(\varphi) = \varnothing$, then (CP) justifies to write $\Gamma(\varphi)$ instead of $\Gamma(\varphi, \gamma)$.

*Let $|.| : M \to \{1, 0, B, N\}$ be the function defined by $|m| = 1 \Leftrightarrow m \in TRUE \smallsetminus FALSE$, $|m| = 0 \Leftrightarrow m \in FALSE \smallsetminus TRUE$, $|m| = B \Leftrightarrow m \in TRUE \cap FALSE$, $|m| = N \Leftrightarrow m \notin TRUE \cup FALSE$, for every proposition $m \in M$. The Gamma-function satisfies the following truth conditions. For all $\varphi, \psi \in Expr(\mathcal{L}, C)$, all $a \in Expr_\mathcal{L}$ and all assignments $\gamma \in M^V$:*

(i) $|\Gamma(\varphi : true, \gamma)| = |\Gamma(\varphi, \gamma)|$

(ii) $\Gamma(\varphi : false, \gamma) \in TRUE \Leftrightarrow \Gamma(\varphi, \gamma) \in FALSE$

(iii) $\Gamma(\varphi : false, \gamma) \in FALSE \Leftrightarrow \Gamma(\varphi, \gamma) \in TRUE$

(iv) $|\Gamma(\varphi \vee \psi, \gamma)| = sup\{|\Gamma(\varphi, \gamma)|, |\Gamma(\psi, \gamma)|\}$

(v) $|\Gamma(\varphi \wedge \psi, \gamma)| = inf\{|\Gamma(\varphi, \gamma)|, |\Gamma(\psi, \gamma)|\}$

(vi) $\Gamma(\varphi \equiv \psi, \gamma) \in FALSE \smallsetminus TRUE \Leftrightarrow \Gamma(\varphi, \gamma) \neq \Gamma(\psi, \gamma)$

(vii) $\Gamma(\varphi < \psi, \gamma) \in FALSE \smallsetminus TRUE \Leftrightarrow \Gamma(\varphi, \gamma) \not<^\mathcal{M} \Gamma(\psi, \gamma)$

(viii) $|\Gamma(\exists x.\varphi, \gamma)| = sup\{|\Gamma(\varphi, \gamma_x^m)| \mid m \in M\}$

(ix) $|\Gamma(\forall x.\varphi, \gamma)| = inf\{|\Gamma(\varphi, \gamma_x^m)| \mid m \in M\}$

(x) $\Gamma(a) \in TRUE \Leftrightarrow a \in T$; and
$\Gamma(a) \in FALSE \Leftrightarrow\, \sim a \in T$ *(Bridge Property).*

There are exactly two one-element models. If the underlying theory $T$ is empty, then the associated one-element model has the property $TRUE = FALSE = \varnothing$, $|M| = 1$, and the model satisfies no formula. On the other hand, if $T = Expr_\mathcal{L}$ is the underlying theory, then the associated one-element model has the property $M = TRUE = FALSE$ and it satisfies all formulas. In both models, the respective reference relation is given by $M \times M$, i.e. a set of exactly one tuple. One easily checks that there cannot exist further one-element models. A classical model is a model where $M$ is the disjoint union of $TRUE$ and $FALSE$. Note that in this classical case the truth conditions specialize to the usual classical truth conditions (see [18]). Observe that truth condition (i) guarantees that the Tarski biconditionals hold, and (x) establishes a "bridge" between the underlying parameter logic and the extension ensuring that truth and falsity w.r.t. the a given theory $T$ of the parameter logic are preserved in the model over $T$ of the $\in_T$-extension.

**Definition 5.2** *Let $\mathcal{M} = (M, TRUE, FALSE, <^\mathcal{M}, \Gamma, T)$ be a model, $\gamma \in M^V$ and $\varphi \in Expr(\mathcal{L}, C)$. The satisfaction relation $\vDash$ is defined by:*

$$(\mathcal{M}, \gamma) \vDash \varphi :\Leftrightarrow \Gamma(\varphi, \gamma) \in TRUE.$$

*The tupel $(\mathcal{M}, \gamma)$ is called an interpretation. If $(\mathcal{M}, \gamma) \vDash \varphi$, then we say that $(\mathcal{M}, \gamma)$ is a model of $\varphi$. If $\varphi$ is a sentence, then we may omit assignments writing $\mathcal{M} \vDash \varphi$. Analogously for sets $\Phi$ of expressions. The consequence relation $\Vdash$ is given in the usual model-theoretical way:*

$$\Phi \Vdash \varphi :\Leftrightarrow \text{ every model of } \Phi \text{ is a model of } \varphi.$$

The following Substitution Principle is guaranteed by the Substitution Property SP (see [18]).

**Lemma 5.3 (Substitution Principle)** *For all formulas $\varphi, \psi, \psi' \in Expr(\mathcal{L}, C)$ and all $x \in V$,*
$$\Vdash \psi \equiv \psi' \to \varphi[x := \psi] \equiv \varphi[x := \psi']$$

**Definition 5.4** *We say that a model $\mathcal{M}$ is*

- *$K_3$ if $TRUE \cap FALSE = \emptyset$,*
- *$P_3$ if $M = TRUE \cup FALSE$,*
- *classical if $\mathcal{M}$ is both $K_3$ and $P_3$,*
- *$B_4$ if there are no constraints at all (i.e., $\mathcal{M}$ may be $K_3$ or $P_3$ or classical or it may be neither $K_3$ nor $P_3$).*

According to this definition, all models are $B_4$ and some are in addition $K_3$, $P_3$ or classical. Note that we have ensured in the model definition that $\mathcal{M}$ is $K_3$, $P_3$, classical iff the underlying complete $B_4(\mathcal{L})$-theory is a $K_3$-, a $P_3$-, a maximal $\mathcal{L}$-theory, respectively. The set of theories of all interpretations $(\mathcal{M}, \gamma)$ generates an abstract logic $B_4(\mathcal{L})^*_C$. If we consider only interpretations $(\mathcal{M}, \gamma)$ where $\mathcal{M}$ is a classical model, then we get an abstract logic denoted by $\mathcal{L}^*_C$. In the next section we will show by means of a complete sequent calculus that this logic has a compact consequence relation. It follows that it is a *classical* abstract logic. If we consider only interpretations $(\mathcal{M}, \gamma)$ where $\mathcal{M}$ is a $K_3$ model, then we get an abstract logic $K_3(\mathcal{L})^*_C$, and similarly for $P_3(\mathcal{L})^*_C$. We leave it here as a claim that $B_4(\mathcal{L})^*_C = B_4(\mathcal{L}^*_C)$, $K_3(\mathcal{L})^*_C = K_3(\mathcal{L}^*_C)$ and $P_3(\mathcal{L})^*_C = P_3(\mathcal{L}^*_C)$. Thus, $B_4(\mathcal{L})^*_C$, $K_3(\mathcal{L})^*_C$ and $P_3(\mathcal{L})^*_C$ are $B_4$, $K_3$, $P_3$ abstract logics, respectively, associated to the ambient classical abstract logic $\mathcal{L}^*_C$, in the sense of Definition 3.1.

**Definition 5.5** *For a given classical abstract logic $\mathcal{L}$ and a set of constant symbols $C$ we say that the abstract logic $\mathcal{L}^*_C$, $B_4(\mathcal{L})^*_C$, $K_3(\mathcal{L})^*_C$, $P_3(\mathcal{L})^*_C$ is the $\in_T$-style (non-Fregean) extension of the parameter logic $\mathcal{L}$, $B_4(\mathcal{L})$, $K_3(\mathcal{L})$, $P_3(\mathcal{L})$, respectively.*

It remains to show that for any given complete $B_4(\mathcal{L})$-theory $T$ there exists a model over $T$. That is, we must ensure that the defined $\in_T$-style extensions actually exist. A standard model is a model where every element of the propositional universe is denoted by a sentence (there are no "non-standard" elements) and where the $<$-connective is interpreted in accordance with its intended meaning, i.e. the following condition of $<$-intensionality is satisfied: For all formulas $\varphi, \psi$ and all assignments $\gamma : V \to M$, if $(\mathcal{M}, \gamma) \vDash \varphi < \psi$, then there are formulas $\varphi'$ and $\psi'$ such that $\varphi' \prec \psi'$ and $\Gamma(\varphi, \gamma) = \Gamma(\varphi', \gamma)$ and $\Gamma(\psi, \gamma) = \Gamma(\psi', \gamma)$ (see [18]). We will show the existence of such models in the following.

As in classical $\in_T$-Logic, we call an interpretation extensional if any two formulas with the same truth value denote the same proposition. The simplest extensional models of classical $\in_T$-Logic are two-element models. In the following we construct an extensional model for our 4-valued logic, which is neither $K_3$ nor $P_3$. The universe $M$ of this model is given by the De Morgan lattice of the 4 truth values $1, 0, B, N$. Let $\mathcal{L}$ be a classical abstract logic and let $T$ be a complete $B_4(\mathcal{L})$-theory and $\overline{T}$ its complement. We suppose that $T$ is neither a $K_3$- nor a $P_3$-theory, i.e. $T \cap \overline{T} \neq \emptyset$ and $T \cup \overline{T} \neq Expr_\mathcal{L}$. We define $M = \{1, 0, B, N\}$, and $TRUE = \{1, B\}$, $FALSE = \{0, B\}$. The reference relation is defined by $<^M = M \times M$. Suppose there is a partition $C_1 \cup C_0 \cup C_B \cup C_N$ on the given set $C$ of constant symbols. The Gamma-function is defined simultaneously for all assignments $\gamma : V \to M$ in the following way:

$$\Gamma(x, \gamma) = \gamma(x), \text{ for } x \in V$$

$$\Gamma(c) = \begin{cases} 1, & \text{if } c \in C_1 \\ 0, & \text{if } c \in C_0 \\ B, & \text{if } c \in C_B \\ N, & \text{if } c \in C_N \end{cases}$$

$$\Gamma(a) = \begin{cases} 1, & \text{if } a \in T \smallsetminus \overline{T} \\ 0, & \text{if } a \in \overline{T} \smallsetminus T \\ B, & \text{if } a \in T \cap \overline{T} \\ N, & \text{if } a \in Expr_\mathcal{L} \smallsetminus (T \cup \overline{T}) \end{cases}$$

$$\Gamma(\varphi : true, \gamma) = \Gamma(\varphi, \gamma)$$

$$\Gamma(\varphi : false, \gamma) = \begin{cases} 1, & \text{if } \Gamma(\varphi, \gamma) = 0 \\ 0, & \text{if } \Gamma(\varphi, \gamma) = 1 \\ B, & \text{if } \Gamma(\varphi, \gamma) = B \\ N, & \text{if } \Gamma(\varphi, \gamma) = N \end{cases}$$

$$\Gamma(\varphi \vee \psi, \gamma) = sup\{\Gamma(\varphi, \gamma), \Gamma(\psi, \gamma)\}$$
$$\Gamma(\varphi \wedge \psi, \gamma) = inf\{\Gamma(\varphi, \gamma), \Gamma(\psi, \gamma)\}$$

$$\Gamma(\varphi \equiv \psi, \gamma) = \begin{cases} 1, & \text{if } \Gamma(\varphi, \gamma) = \Gamma(\psi, \gamma) \\ 0, & \text{else} \end{cases}$$

$$\Gamma(\varphi < \psi, \gamma) = 1$$

$$\Gamma(\exists x.\varphi, \gamma) = sup\{\Gamma(\varphi, \gamma_x^m) \mid m \in M\}$$
$$\Gamma(\forall x.\varphi, \gamma) = inf\{\Gamma(\varphi, \gamma_x^m) \mid m \in M\}$$

It is clear that the Gamma-function satisfies the truth conditions. (EP) and (RP) are trivially satisfied. (CP) and (SP) follow by induction on the expressions. We show only a quantifier case of (SP) and leave the remaining cases to the reader. Let $\varphi = \exists x.\psi$, and let $\sigma : V \to Expr(\mathcal{L}, C)$ be a substitution. First, we show that for any $m \in M$ and all $y \in fvar(\psi)$ the following holds:

(5.1) $$(\gamma\sigma)_x^m(y) = (\gamma_z^m \sigma[x := z])(y),$$

where $z$ is the variable forced by $\sigma$ w.r.t. $\exists x.\psi$. That is, $z := lub(fvar((\exists x.\psi)[\sigma])$ and $\varphi[\sigma] = (\exists x.\psi)[\sigma] = \exists z.\psi[\sigma[x := z]]$. Let $y \in fvar(\psi)$. First, suppose $y = x$. Then $(\gamma\sigma)_x^m(y) = m$. On the other hand, $(\gamma_z^m \sigma[x := z])(y) = \Gamma(\sigma[x := z](y), \gamma_z^m) = \Gamma(z, \gamma_z^m) = \gamma_z^m(z) = m$. Now suppose that $y \neq x$. Note that by definition, $z \notin fvar(\sigma(y))$. Then by (CP) we get $(\gamma\sigma)_x^m(y) = (\gamma\sigma)(y) = \Gamma(\sigma(y), \gamma) = \Gamma(\sigma(y)\gamma_z^m) = \Gamma(\sigma[x := z](y)\gamma_z^m) = (\gamma_z^m \sigma[x := z])(y)$. Consequently:

$$|\Gamma(\exists x.\psi, \gamma\sigma)| = sup\{|\Gamma(\psi, (\gamma\sigma)_x^m)| \mid m \in M\}$$
$$= sup\{|\Gamma(\psi, \gamma_z^m \sigma[x := z])| \mid m \in M\}, \text{ by (5.1) and (CP)}$$
$$= sup\{|\Gamma(\psi[\sigma[x := z]], \gamma_z^m)| \mid m \in M\}, \text{ by induction hypothesis}$$
$$= |\Gamma(\exists z.\psi[\sigma[x := z]], \gamma_z^m)|$$
$$= |\Gamma((\exists x.\psi)[\sigma], \gamma)|$$

Thus, $\mathcal{M} = (M, TRUE, FALSE, <^{\mathcal{M}}, \Gamma, T)$ is an extensional model. It remains to show that $\mathcal{M}$ is a standard model. Since $T$ is neither a complete $K_3$- nor a complete $P_3$-theory one recognizes that for every proposition $m \in M$ there is a sentence $\varphi$ such that $\Gamma(\varphi) = m$. Thus, there are no non-standard elements. We show that $\mathcal{M}$ is $<$-intensional. Let $(\mathcal{M}, \gamma) \models \varphi < \psi$. In fact, this is true for all formulas $\varphi, \psi$. Suppose $\Gamma(\varphi, \gamma) = 1$ and $\Gamma(\psi, \gamma) = N$. Consider the formulas $x \equiv x$ and $(x \equiv x) \wedge b$, where $b$ is an $\mathcal{L}$-formula with $b \notin T \cup \overline{T}$. Clearly, $x \equiv x \prec (x \equiv x) \wedge b$ and $\Gamma(x \equiv x, \gamma) = 1$ and $\Gamma(b, \gamma) = N$. The other cases can be shown in a similar way. It follows that $\mathcal{M}$ is a standard model. Similarly, one shows that for any given complete $\mathcal{L}$-theory (complete $K_3(\mathcal{L})$-theory, complete $P_3(\mathcal{L})$-theory) $T$ there is an extensional standard model over $T$.

We have shown the following

**Theorem 5.6** *Let $\mathcal{L}$ be a classical abstract logic. For every complete $B_4(\mathcal{L})$-theory $T$ there exists a standard model over $T$.*

The following definition of *extension* is inspired by a similar concept given in [29].

**Definition 5.7** *Let $\mathcal{L}, \mathcal{L}'$ be minimally generated abstract logics.*[18]

- *$\mathcal{L}'$ is a (conservative) extension of $\mathcal{L}$, notation: $\mathcal{L} \leq \mathcal{L}'$, if $Expr_{\mathcal{L}} \subseteq Expr_{\mathcal{L}'}$ and $Th_{\mathcal{L}} = \{T' \cap Expr_{\mathcal{L}} \mid T' \in Th_{\mathcal{L}'}\}$.*

- *$\mathcal{L}'$ is (in an abstract model-theoretic sense) a sublogic of $\mathcal{L}$, notation: $\mathcal{L}' \subseteq \mathcal{L}$, if $Expr_{\mathcal{L}} = Expr_{\mathcal{L}'}$ and $TPTh_{\mathcal{L}'} \subseteq TPTh_{\mathcal{L}}$. If $\mathcal{L}' \subseteq \mathcal{L}$, then we call $\mathcal{L}$ a superlogic of $\mathcal{L}'$.*[19]

**Remark 5.8**  • *In the literature, a logic is often identified with its set of theorems and the concept of sublogic is given as the relation of inclusion w.r.t. the corresponding sets of theorems. Note, however, that this current notion of sublogic is not always accurate. For instance, Priest's Logic of Paradox $P_3$ and classical propositional logic have the same set of theorems. This follows readily from the fact that every complete theory of classical logic is a complete $P_3$-theory, and on the other hand, every complete $P_3$-theory contains a complete theory of classical logic. So it seems to be better to identify a logic with its set of theories.*

---

[18]Recall that $TPTh_{\mathcal{L}}$ and $TPTh_{\mathcal{L}'}$ are the respective minimal generator sets.

[19]Recall that the totally prime theories represent, in an abstract way, models of the logic. In this sense, the set of complete theories represents the set of all models.

- *Let $\mathcal{L}_I$ be an intuitionistic abstract logic which is not classical. We saw that the set of complete theories is precisely the set of prime theories, and there are prime theories which are not maximal. The set of maximal $\mathcal{L}_I$-theories generates a sublogic $\mathcal{L}_{Cl} \subseteq \mathcal{L}_I$, which is a classical abstract logic. Furthermore, $\mathcal{L}_{Cl} \subseteq K_3(\mathcal{L}_{Cl}) \subseteq B_4(\mathcal{L})$ and $\mathcal{L}_{Cl} \subseteq P_3(\mathcal{L}_{Cl}) \subseteq B_4(\mathcal{L})$. In this partial ordering of abstract logics, the logic $\mathcal{L}_I$ is incomparable with $K_3(\mathcal{L}_{Cl})$, $P_3(\mathcal{L}_{Cl})$ and $B_4(\mathcal{L}_{Cl})$.*

The model existence Theorem 5.6 implies that a $\in_T$-style extension is a conservative extension of its underlying parameter logic in the sense of Definition 5.7:

**Corollary 5.9** *Let $\mathcal{L}$ be a classical abstract logic and $C$ a set of constant symbols. Then $\mathcal{L} \leq \mathcal{L}_C^*$, $K_3(\mathcal{L}) \leq K_3(\mathcal{L}_C^*)$, $P_3(\mathcal{L}) \leq P_3(\mathcal{L}_C^*)$ and $B_4(\mathcal{L}) \leq B_4(\mathcal{L}_C^*)$.*

The following fact will be useful. The proof is an easy exercise.

**Lemma 5.10** *Suppose $\mathcal{L} \leq \mathcal{L}'$ and $A \cup \{a\} \subseteq Expr_\mathcal{L}$. Then $A \Vdash_\mathcal{L} a \Leftrightarrow A \Vdash_{\mathcal{L}'} a$.*

# 6 A sequent calculus for the classical case

Suppose we are given a set $C$ of constant symbols and a classical abstract logic $\mathcal{L}$ with a sound and complete sequent calculus $\mathcal{K}$. The notion of derivation in $\mathcal{K}$ is defined as usual. A set $A$ of $\mathcal{L}$-expressions is said to be $\mathcal{K}$-consistent if there is some $\mathcal{L}$-expression $b$ such that $b$ is not derivable from $A$. Otherwise, $A$ is $\mathcal{K}$-inconsistent. A set $A$ of expressions is maximally $\mathcal{K}$-consistent if $A$ is $\mathcal{K}$-consistent and no proper extension of $A$ is $\mathcal{K}$-consistent.

The aim of this section is to show that under these assumptions we can find a sequent calculus which contains the rules of $\mathcal{K}$ and is sound and complete with respect to the $\in_T$-style extension $\mathcal{L}_C^*$.[20] Consequently, the $\in_T$-style extension $\mathcal{L}_C^*$ is compact. Thus, it is a classical abstract logic, too. Besides the rules of $\mathcal{K}$ the extended calculus will contain pure $\in_T$-rules as well as bridge rules which reflect the semantic Bridge Property. It turns out that the actual form of the rules of $\mathcal{K}$ is not relevant. In the following, we list only the pure $\in_T$-rules and the bridge rules assuming that the rules of $\mathcal{K}$ are given. The system, denoted by $\mathcal{K}^*_C$, is a version of Sträter's original sequent calculus [25] improved in some aspects and extended by rules concerning the new reference connective $<$ and by the bridge rules (R17) and (R18). We use $\varphi \to \psi$ as an abbreviation for $\varphi : false \vee \psi$.

---

[20] A sound and complete sequent calculus for pure $\in_T$-Logic was presented by Sträter [25]. Sträter's construction was simplified and extended by Lewitzka [11] where the $\in_T$-extension of classical first-order logic is defined and a sound and complete sequent calculus for the extension (with similar Bridge Rules as given in the present paper) was developed. A more general result, based on a Hilbert-style calculus, is given by Zeitz [29] who essentially shows that the sound and complete Hilbert-style axiomatization of a given classical abstract logic can be extended to such an axiomatization for the corresponding $\in_T$-extension. We show here a version of Zeitz's result working with a sequent calculus instead of Hilbert-style calculus. This section represents essentially an amalgam of revised, simplified and improved constructions and technical machinery originally developed in [25, 29, 11].

(R1) $\dfrac{}{\Delta \vdash \varphi}$ if $\varphi \in \Delta$ 
(R2) $\dfrac{\Delta \vdash \varphi}{\Delta' \vdash \varphi}$ if $\Delta \subseteq \Delta'$

(R3) $\dfrac{\Delta \vdash \varphi, \Delta \vdash \varphi : false}{\Delta \vdash \psi}$ 
(R4) $\dfrac{\Delta \cup \{\varphi\} \vdash \psi, \Delta \cup \{\varphi : false\} \vdash \psi}{\Delta \vdash \psi}$

(R5) $\dfrac{\Delta \vdash \varphi}{\Delta \vdash \varphi \vee \psi}$ 
(R6) $\dfrac{\Delta \vdash \varphi}{\Delta \vdash \psi \vee \varphi}$

(R7) $\dfrac{\Delta \cup \{\varphi_1\} \vdash \psi, \Delta \cup \{\varphi_2\} \vdash \psi}{\Delta \cup \{\varphi_1 \vee \varphi_2\} \vdash \psi}$

(R8) $\dfrac{\Delta \vdash \varphi[x := \psi]}{\Delta \vdash \exists z.(\varphi[x := z])}$ if $x \in fvar(\varphi)$ and $z \notin fvar(\varphi) \smallsetminus \{x\}$

(R9) $\dfrac{\Delta \cup \{\varphi[x := y]\} \vdash \psi}{\Delta \cup \{\exists z.(\varphi[x := z])\} \vdash \psi}$ if $x \in fvar(\varphi)$, $z \notin fvar(\varphi) \smallsetminus \{x\}$ and
$y \notin fvar(\Delta, \exists x.\varphi, \psi)$

(R10) $\dfrac{\Delta \vdash \psi \equiv \psi'}{\Delta \vdash \varphi[x := \psi] \equiv \varphi[x := \psi']}$ 
(R11) $\dfrac{\Delta \vdash \psi \equiv \psi'}{\Delta \vdash \psi \to \psi'}$

(R12) $\dfrac{}{\Delta \vdash \varphi \equiv \varphi'}$ if $\varphi =_\alpha \varphi'$ 
(R13) $\dfrac{}{\Delta \vdash \varphi < \psi}$ if $\varphi \prec \psi$

(R14) $\dfrac{\Delta \vdash \varphi < \psi, \Delta \vdash \psi < \chi}{\Delta \vdash \varphi < \chi}$

(R15) $\dfrac{\Delta \vdash \varphi}{\Delta \vdash \varphi : true}$ 
(R16) $\dfrac{\Delta \vdash \varphi : true}{\Delta \vdash \varphi}$

(R17) $\dfrac{\Delta \vdash \sim a}{\Delta \vdash a : false}$ where $a \in Expr_{\mathcal{L}}$

(R18) $\dfrac{\Delta \vdash a : false}{\Delta \vdash \sim a}$ where $a \in Expr_{\mathcal{L}}$

For $\Phi \cup \{\varphi\} \subseteq Expr(\mathcal{L}, C)$ we write $\Phi \vdash_{\mathcal{K}_C^*} \varphi$ if there is a finite $\Delta \subseteq \Phi$ and a derivation of the sequent $\Delta \vdash \varphi$ in calculus $\mathcal{K}_C^*$. The following result shows that the notion of derivation is in some sense independent of the given set of constant symbols $C$. A similar result is shown in [25]. It is enough to prove that the rules are invariant under substitutions. We leave this as an exercise to the reader.

**Lemma 6.1** *If $(\Delta_i \vdash \varphi_i)_{i \leq n}$ is a derivation in $\mathcal{K}_C^*$, $c \in C$ and $x_c \in V \smallsetminus \bigcup_{i \leq n} var(\Delta_i \cup \{\varphi_i\})$, then $(\Delta_i[c := x_c] \vdash \varphi_i[c := x_c])_{i \leq n}$ is also a derivation.*[21]

**Corollary 6.2** *The notion of derivation is independent of the underlying set of constant symbols. More precisely, if $C' := con(\Phi \cup \{\varphi\})$ and $C' \subseteq C''$, then $\Phi \vdash_{\mathcal{K}_{C'}^*} \varphi \Leftrightarrow \Phi \vdash_{\mathcal{K}_{C''}^*} \varphi$.*

In the following we may omit the index $\mathcal{K}_C^*$ writing $\Phi \vdash \varphi$ in order to express that $\varphi$ is derivable from $\Phi$ in $\mathcal{K}_C^*$. Moreover, we may assume that $C$ is the set of all constant symbols occurring in $\Phi$ and $\varphi$.

### 6.1 Soundness and Completeness

It is straightforward to prove that every rule of the extension is sound.

**Theorem 6.3 (Soundness)** *For all $\Phi \cup \{\varphi\} \subseteq Expr(\mathcal{L}, C)$, the following holds: $\Phi \vdash_{\mathcal{K}_C^*} \varphi \Rightarrow \Phi \Vdash_{\mathcal{K}_C^*} \varphi$.*

The notions of $\mathcal{K}_C^*$-consistency and maximal $\mathcal{K}_C^*$-consistency are defined in the usual way. In order to prove the Completeness Theorem we define a notion of Henkin set and show that every maximally $\mathcal{K}_C^*$-consistent Henkin set has a model. Showing that every $\mathcal{K}_C^*$-consistent set extends to a maximally $\mathcal{K}_C^*$-consistent Henkin set will complete the proof of the Completeness Theorem.

**Lemma 6.4** *Let $\Phi \subseteq Expr(\mathcal{L}, C)$.*

---

[21]For $\Delta$ a set of expressions and $\sigma$ a substitution we write $\Delta[\sigma]$ for the set $\{\psi[\sigma] \mid \psi \in \Delta\}$. $var(\Delta)$ denotes the set of all variables occurring in $\Delta$.

(i) $\Phi$ is $\mathcal{K}_C^*$-inconsistent if and only if $\Phi \vdash \varphi$ and $\Phi \vdash \neg\varphi$, for some expression $\varphi \in Expr(\mathcal{L}, C)$.

(ii) $\Phi$ is maximally $\mathcal{K}_C^*$-consistent if and only if $\Phi$ is $\mathcal{K}_C^*$-consistent and for all expressions $\varphi \in Expr(\mathcal{L}, C)$, $\varphi \in \Phi$ or $\neg\varphi \in \Phi$.

The facts of the next lemma follow from standard arguments or derive easily from corresponding rules of the calculus.

**Lemma 6.5** *Let $\Phi$ be a maximally $\mathcal{K}_C^*$-consistent set of expressions. Then for all expressions $\varphi, \psi \in Expr(\mathcal{L}, C)$:*

(i) $\varphi \in \Phi \Leftrightarrow \Phi \vdash \varphi$.

(ii) $\varphi : false \in \Phi \Leftrightarrow \varphi \notin \Phi$.

(iii) $\varphi : true \in \Phi \Leftrightarrow \varphi \in \Phi$.

(iv) $\varphi \vee \psi \in \Phi \Leftrightarrow \varphi \in \Phi$ or $\psi \in \Phi$.

(v) *If $\chi$ is a formula such that $x \in fvar(\chi)$ and $\chi[x := \psi] \in \Phi$, then $\exists x.\chi \in \Phi$.*

(vi) *If $\varphi =_\alpha \psi$, then $\varphi \equiv \psi \in \Phi$.*

(vii) *If $\varphi \prec \psi$, then $\varphi < \psi \in \Phi$.*

(viii) *$\Phi$ is closed under Modus Ponens, i.e., if $\varphi \in \Phi$ and $\varphi \to \psi \in \Phi$, then $\psi \in \Phi$.*[22]

**Definition 6.6** *For $\Phi \subseteq Expr(\mathcal{L}, C)$ we define $\mathcal{L}(\Phi) := \Phi \cap Expr_{\mathcal{L}}$.*

A version of the following Proposition was first shown in [11].

**Proposition 6.7** *If $\Phi$ is a $\mathcal{K}_C^*$-consistent set, then $\mathcal{L}(\Phi)$ is $\mathcal{K}$-consistent. If the set $\Phi$ is maximally $\mathcal{K}^*$-consistent, then $\mathcal{L}(\Phi)$ is a maximal $\mathcal{L}$-theory.*

---

[22]Recall that $\varphi \to \psi$ is defined as $\varphi : false \vee \psi$.

**Proof.** Let $\Phi \subseteq Expr(\mathcal{L}, C)$ be $\mathcal{K}_C^*$-consistent and suppose – towards a contradiction – that $\mathcal{L}(\Phi)$ is $\mathcal{K}$-inconsistent. Then every $a \in Expr_\mathcal{L}$ is derivable from $\mathcal{L}(\Phi)$ in $\mathcal{K}$ and therefore also in the extended calculus $\mathcal{K}_C^*$. In particular, $a$ and $\sim a$ are derivable. By bridge rule (R17) and rule (R3), $\Phi$ is $\mathcal{K}_C^*$-inconsistent – a contradiction. Now let $\Phi$ be maximally $\mathcal{K}_C^*$-consistent. In order to prove that $\mathcal{L}(\Phi)$ is a $\mathcal{L}$-theory it is enough to show that $\mathcal{L}(\Phi)$ is deductively closed (with respect to $\Vdash_\mathcal{L}$) and $\mathcal{L}$-consistent. We first show that $\mathcal{L}(\Phi)$ is deductively closed in logic $\mathcal{L}$. Suppose $\mathcal{L}(\Phi) \Vdash_\mathcal{L} b$, for some $b \in Expr_\mathcal{L}$. By the completeness theorem of $\mathcal{L}$, $\mathcal{L}(\Phi) \vdash_\mathcal{K} b$. That is, $b$ is derivable from $\mathcal{L}(\Phi)$ in calculus $\mathcal{K}$. Then $b$ is derivable from $\Phi$ in calculus $\mathcal{K}_C^*$. By (i) of Lemma 6.5, $b \in \Phi$. But this means that $b \in \mathcal{L}(\Phi)$ and $\mathcal{L}(\Phi)$ is deductively closed. Since $\mathcal{L}(\Phi)$ is $\mathcal{K}$-consistent and the logic $\mathcal{L}$ is regular, $\mathcal{L}(\Phi)$ is consistent in $\mathcal{L}$ (i.e., it is contained in some $\mathcal{L}$-theory). Suppose $T = \mathcal{L}(\Phi)$ is not a maximal theory. Then there is a maximal theory $T' \supseteq T$ and some $a \in T' \smallsetminus T$. Since $a \notin \Phi$, Lemma 6.5 yields $a : false \in \Phi$. By the bridge rule (R18), $\sim a \in \Phi$. But this implies $\sim a \in T'$, contradicting $a \in T'$. Thus, $T$ must be a maximal $\mathcal{L}$-theory. $\square$

**Lemma 6.8** *Let $\Phi \subseteq Expr(\mathcal{L}, C)$ be a maximally $\mathcal{K}_C^*$-consistent set. Define $\varphi \approx \psi :\Leftrightarrow \varphi \equiv \psi \in \Phi$. Then the following holds:*

*(i) If $\varphi$ is an expression, $x \in V$ and $\psi \approx \psi'$, then $\varphi[x := \psi] \approx \varphi[x := \psi']$.*

*(ii) Let $\varphi \approx \varphi'$ and $\psi \approx \psi'$. Then $(\varphi < \psi) \approx (\varphi' < \psi')$.*

*(iii) $\approx$ is an equivalence relation containing alpha-congruence.*

**Proof.** (i): This follows applying (R10).
(ii): Consider the expression $(x < y)$ with variables $x \neq y$ and $y \notin fvar(\varphi) \cup fvar(\varphi')$. By (i), $(x < y)[x := \varphi] \approx (x < y)[x := \varphi']$. That is, $(\varphi < y) \approx (\varphi' < y)$. To this we apply the substitutions $[y := \psi]$ and $[y := \psi']$ and get $(\varphi < \psi) \approx (\varphi' < \psi')$.
(iii): By (R12), $\approx$ contains the relation of alpha-congruence and is therefore in particular reflexive. Towards symmetry suppose $\varphi \approx \psi$. Let $\chi := (x \equiv \varphi)$, where $x \notin fvar(\varphi)$. Then $\chi[x := \varphi] \approx \chi[x := \psi]$, by (R10). Since $\chi[x := \varphi] \in \Phi$, we get $\chi[x := \psi] \in \Phi$, by Lemma 6.5. Thus, $\psi \approx \varphi$ and the symmetry of $\approx$ follows. Now suppose $\varphi_1 \approx \varphi_2$ and $\varphi_2 \approx \varphi_3$. Let $\chi := (\varphi_1 \equiv x)$, where variable $x \notin fvar(\varphi)$. Then $\chi[x := \varphi_2] \approx \chi[x := \varphi_3]$, by (R10). Then (R11) and Modus Ponens yield $\chi[x := \varphi_3] \in \Phi$. That is, $\varphi_1 \approx \varphi_3$ and $\approx$ is transitive. $\square$

Item (ii) of the preceding Lemma says that $\approx$ is compatible with the reference connective $<$. Of course, in a similar way one can prove that $\approx$ is compatible with the other connectives and operators, too. In this sense, $\approx$ is a congruence.

**Definition 6.9** $\Phi \subseteq Expr(\mathcal{L}, C)$ *is called a Henkin set if* $\exists x.\psi \in \Phi$ *implies the existence of some* $c \in C$ *such that* $\psi[x := c] \in \Phi$.

**Theorem 6.10** *Every maximally $\mathcal{K}_C^*$-consistent Henkin set has a model.*

**Proof.** Let $\Phi \subseteq Expr(\mathcal{L}, C)$ be a maximally $\mathcal{K}_C^*$-consistent Henkin set. For formulas $\varphi, \psi$ we define $\varphi \approx \psi :\Leftrightarrow \varphi \equiv \psi \in \Phi$. By Lemma 6.8 we know that $\approx$ is an equivalence relation on the set of expressions. The universe is given by the set of equivalence classes of *sentences* modulo $\approx$:

$$M := Sent(\mathcal{L}, C)/_\approx.$$

Notice that we consider here the restriction of $\approx$ to a relation on the set of sentences. For $\varphi \in Sent(\mathcal{L}, C)$ let $\overline{\varphi} = \{\psi \in Sent(\mathcal{L}, C) \mid \varphi \approx \psi\}$. We define the relation $<^\mathcal{M}$ on $M$ by

$$\overline{\varphi} <^\mathcal{M} \overline{\psi} :\Leftrightarrow \varphi < \psi \in \Phi.$$

Let us show that $<^\mathcal{M}$ is well-defined. Suppose $\varphi < \psi \in \Phi$ and $\varphi', \psi'$ are sentences such that $\varphi \approx \varphi'$ and $\psi \approx \psi'$. Then by Lemma 6.8, $\varphi < \psi \approx \varphi' < \psi'$. By Lemma 6.5, $\varphi' < \psi' \in \Phi$. Thus, $<^\mathcal{M}$ is well-defined.

For each assignment $\beta : V \to M$ let $\tau_\beta : V \to Sent(\mathcal{L}, C)$ be a function with the property $\tau_\beta(x) \in \beta(x)$ for all $x \in V$. We define the Gamma-function by

$$\Gamma(\varphi, \beta) = \overline{c} :\Leftrightarrow c \approx \varphi[\tau_\beta].$$

Note that the function $\tau_\beta$ is both an assignment and a substitution. We know that $\varphi[\tau_\beta]$ is the result of the *simultaneous* replacements of all free variables in $\varphi$ by *sentences*. Since we deal with sentences, these replacements can be carried out successively. That is, one can split $\tau_\beta$ into a series of substitutions of the form $[x := \psi_i]$, for sentences $\psi_i \in \beta(x)$. Thus, by the first item of Lemma 6.8, it follows that the result $\varphi[\tau_\beta]$ is independ of the choice $\tau_\beta(x) \in \beta(x)$. Furthermore, for each formula $\psi$ there is a constant symbol $c \in C$ such that $c \equiv \psi \in \Phi$. This can be seen as follows. Consider the expression $(x \equiv \psi)$, where $x \in V \smallsetminus fvar(\psi)$, and the derivation

$$\varnothing \vdash (x \equiv \psi)[x := \psi], \text{ (R12)}$$
$$\varnothing \vdash \exists x.(x \equiv \psi), \text{ (R8)}$$

Since $\Phi$ is a maximally $\mathcal{K}_C^*$-consistent Henkin set, there is some $c \in C$ such that $c \equiv \psi \in \Phi$. We conclude that the Gamma-function is well-defined. Now we put

$$TRUE := \{\overline{c} \mid c \in \Phi\},$$
$$FALSE := \{\overline{c} \mid c : false \in \Phi\} = \{\overline{c} \mid c \notin \Phi\}.$$

By Lemma 6.5, the set $TRUE$ is well-defined. Let $(c : false) \in \Phi$ and $c \approx c'$. Consider $\psi := (x : false)$. By Lemma 6.8, $\psi[x := c] \approx \psi[x := c']$, i.e. $(c : false) \approx (c' : false)$. By Lemma 6.5, $(c' : false) \in \Phi$. Thus, $FALSE$ is well-defined, too. From the definition of the Gamma-function it follows that

(6.1) $$\Gamma(\varphi, \beta) = \overline{\varphi[\tau_\beta]}.$$

We prove that $\mathcal{M} = (M, TRUE, FALSE, <^{\mathcal{M}}, \Gamma, T)$ is a model, where $T := \mathcal{L}(\Phi)$ is a maximal $\mathcal{L}$-theory, by Proposition 6.7. EP follows immediately. Let $\varphi$ be an expression and suppose $\beta(x) = \beta'(x)$ for all $x \in fvar(\varphi)$. Then $\tau_\beta(x) \approx \tau_{\beta'}(x)$ for all $x \in fvar(\varphi)$. Lemma 6.8 yields $\varphi[\tau_\beta] \approx \varphi[\tau_{\beta'}]$, hence CP holds true. In order to prove the SP suppose that $\sigma : V \to Expr(\mathcal{L}, C)$ is a substitution and $\varphi$ is any expression. We must show: $\Gamma(\varphi[\sigma], \beta) = \overline{\varphi[\sigma][\tau_\beta]} = \overline{\varphi[\tau_{\beta\sigma}]} = \Gamma(\varphi, \beta\sigma)$. We know that $\varphi[\sigma][\tau_\beta] = \varphi[\sigma \circ \tau_\beta]$. Thus, it is enough to show

$$\varphi[\sigma \circ \tau_\beta] \approx \varphi[\tau_{\beta\sigma}],$$

for all $\beta \in M^V$. Let $x \in fvar(\varphi)$. Then on the one hand $(\sigma \circ \tau_\beta)(x) = \sigma(x)[\tau_\beta]$. And on the other hand, $\tau_{\beta\sigma}(x) \in \beta\sigma(x) = \Gamma(\sigma(x), \beta) = \overline{\sigma(x)[\tau_\beta]}$. Hence, $(\sigma \circ \tau_\beta)(x) \approx \tau_{\beta\sigma}(x)$, for all $x \in fvar(\varphi)$. Now, Lemma 6.8 yields the assertion. Thus, SP holds. In order to prove RP let $\varphi, \psi$ be expressions such that $\varphi \prec \psi$. Let $\beta$ be any assignment. We know that $\varphi[\tau_\beta] \prec \psi[\tau_\beta]$. By Lemma 6.5, $\varphi[\tau_\beta] < \psi[\tau_\beta] \in \Phi$. Thus, $\overline{\varphi[\tau_\beta]} <^{\mathcal{M}} \overline{\psi[\tau_\beta]}$. That is, $\Gamma(\varphi, \beta) <^{\mathcal{M}} \Gamma(\psi, \beta)$. Thus, RP holds. Now let $\vartheta : Var \to M$ be the assignment defined by

$$x \mapsto \overline{c}, \text{ if } c \equiv x \in \Phi.$$

Recall that we have already shown that for every $x \in V$ there is some $c \in C$ such that $c \equiv x \in \Phi$. **Claim**:

(6.2) $$\varphi[\tau_\vartheta] \in \Phi \Leftrightarrow \varphi \in \Phi,$$

for all expressions $\varphi$. First, we show $\tau_\vartheta(x) \equiv x \in \Phi$, for all $x \in fvar(\varphi)$. So let $x \in fvar(\varphi)$. Then $\tau_\vartheta(x) \in \vartheta(x) = \overline{c}$, for $c \equiv x \in \Phi$. That is, $\tau_\vartheta(x) \approx c$. $c \approx x$ and transitivity yield $\tau_\vartheta(x) \equiv x \in \Phi$. Thus, $\tau_\vartheta(x) \equiv \varepsilon(x) \in \Phi$, where $\varepsilon$ is the identity substitution $x \mapsto x$. Applying successively (R10) for each $x \in fvar(\varphi)$, we conclude that $\varphi[\tau_\vartheta] \equiv \varphi[\varepsilon] \in \Phi$. Finally, by symmetry, (R11) and Modus Ponens, we get $\varphi[\tau_\vartheta] \in \Phi \Leftrightarrow \varphi[\varepsilon] \in \Phi \Leftrightarrow \varphi \in \Phi$ (recall that $\varphi[\varepsilon] =_\alpha \varphi$ and therefore $\varphi[\varepsilon] \equiv \varphi \in \Phi$, by (R12)). Thus, claim (6.2) holds true. **Claim**:

(6.3) $$\begin{aligned}TRUE &= \{\overline{\varphi[\tau_\vartheta]} \mid \varphi \in \Phi\}, \\ FALSE &= \{\overline{\varphi[\tau_\vartheta]} \mid \varphi : false \in \Phi\} = \{\overline{\varphi[\tau_\vartheta]} \mid \varphi \notin \Phi\}.\end{aligned}$$

By the previous results we have $\varphi \in \Phi \Leftrightarrow \varphi[\tau_\vartheta] \in \Phi \Leftrightarrow c \in \Phi$, for any constant symbol $c$ with $c \equiv \varphi[\tau_\vartheta] \in \Phi$ (as we have seen, such constants exist). This proves claim (6.3). **Claim**:

$$
\begin{aligned}
TRUE &= \{\overline{\varphi} \mid \varphi \in \Phi \cap Sent(\mathcal{L}, C)\}, \\
FALSE &= \{\overline{\varphi} \mid (\varphi : false) \in \Phi \cap Sent(\mathcal{L}, C)\} \\
&= \{\overline{\varphi} \mid \varphi \in Sent(\mathcal{L}, C) \smallsetminus \Phi\}.
\end{aligned}
\tag{6.4}
$$

Suppose that $\varphi \in \Phi$ is a sentence. Then $\varphi = \varphi[\tau_\vartheta]$. Thus, $\overline{\varphi} = \overline{\varphi[\tau_\vartheta]} \in TRUE$, by (6.3). Now suppose $\overline{\varphi} \in TRUE$. By definition of the set $TRUE$ there is some constant symbol $c$ such that $\overline{\varphi} = \overline{c}$ and $c \in \Phi$. By (R11) and Modus Ponens, $\varphi \in \Phi$. Moreover, all elements of a class $\overline{c}$ are sentences. Thus, $\varphi$ is a sentence. The case for the set $FALSE$ follows similarly.

It remains to show that the truth conditions hold. We concentrate on four conditions and the Bridge Property. The remaining cases are left to the reader. Let $\varphi$ be any formula and $\beta : V \to M$ any assignment. Then using (6.1) we get:

$$
\begin{aligned}
\Gamma(\varphi : true, \beta) \in TRUE &\Longleftrightarrow \overline{\varphi : true[\tau_\beta]} \in TRUE \\
&\Longleftrightarrow \overline{\varphi[\tau_\beta] : true} \in TRUE \\
&\stackrel{(6.4)}{\Longleftrightarrow} \varphi[\tau_\beta] : true \in \Phi \\
&\Longleftrightarrow \varphi[\tau_\beta] \in \Phi \\
&\stackrel{(6.4)}{\Longleftrightarrow} \overline{\varphi[\tau_\beta]} \in TRUE \\
&\Longleftrightarrow \Gamma(\varphi, \beta) \in TRUE.
\end{aligned}
$$

$$
\begin{aligned}
\Gamma(\varphi \equiv \psi, \beta) \in TRUE &\Longleftrightarrow \overline{(\varphi \equiv \psi)[\tau_\beta]} \in TRUE \\
&\Longleftrightarrow \overline{\varphi[\tau_\beta] \equiv \psi[\tau_\beta]} \in TRUE \\
&\stackrel{(6.4)}{\Longleftrightarrow} \varphi[\tau_\beta] \equiv \psi[\tau_\beta] \in \Phi \\
&\Longleftrightarrow \varphi[\tau_\beta] \approx \psi[\tau_\beta] \\
&\Longleftrightarrow \overline{\varphi[\tau_\beta]} = \overline{\psi[\tau_\beta]} \\
&\Longleftrightarrow \Gamma(\varphi, \beta) = \Gamma(\psi, \beta)
\end{aligned}
$$

$$
\begin{aligned}
\Gamma(\varphi < \psi, \beta) \in TRUE &\Longleftrightarrow \overline{(\varphi < \psi)[\tau_\beta]} \in TRUE \\
&\Longleftrightarrow \overline{\varphi[\tau_\beta] < \psi[\tau_\beta]} \in TRUE \\
&\stackrel{(6.4)}{\Longleftrightarrow} \varphi[\tau_\beta] < \psi[\tau_\beta] \in \Phi \\
&\Longleftrightarrow \overline{\varphi[\tau_\beta]} <^{\mathcal{M}} \overline{\psi[\tau_\beta]}, \text{ by definition of } <^{\mathcal{M}} \\
&\Longleftrightarrow \Gamma(\varphi, \beta) <^{\mathcal{M}} \Gamma(\psi, \beta)
\end{aligned}
$$

$$\begin{aligned}
\Gamma(\exists x.\varphi, \beta) \in TRUE &\iff \overline{(\exists x.\varphi)[\tau_\beta]} \in TRUE \\
&\iff (\exists x.\varphi)[\tau_\beta] \in \Phi \\
&\iff \exists y.(\varphi[\tau_\beta[x := y]]) \in \Phi \\
&\stackrel{(*)}{\iff} \varphi[\tau_\beta[x := y]][y := c] \in \Phi, \text{ for some } c \in C, \\
&\iff \varphi[\tau_\beta[x := c]] \in \Phi \\
&\stackrel{(**)}{\iff} \varphi[\tau_{\beta_x^{\overline{c}}}] \in \Phi \\
&\iff \overline{\varphi[\tau_{\beta_x^{\overline{c}}}]} \in TRUE \\
&\iff \Gamma(\varphi, \beta_x^{\overline{c}}) \in TRUE,
\end{aligned}$$

where $y$ is the variable forced by the substitution $\tau_\beta$ w.r.t. $\exists x.\varphi$. The equivalence (*) follows from Lemma 6.5 and from the fact that $\Phi$ is a Henkin set. The equivalence (**) is justified as follows: Let $z \in fvar(\varphi)$. First, we suppose $z \neq x$. Then $\tau_\beta[x := c](z) = \tau_\beta(z) \in \beta(z)$. On the other hand, $\tau_{\beta_x^{\overline{c}}}(z) \in \beta_x^{\overline{c}}(z) = \beta(z)$. Thus, $\tau_\beta[x := c](z) \approx \tau_{\beta_x^{\overline{c}}}(z)$. Now suppose $z = x$. Then $\tau_\beta[x := c](z) = c$. On the other hand, $\tau_{\beta_x^{\overline{c}}}(z) \in \beta_x^{\overline{c}}(z) = \overline{c}$. Again, $\tau_\beta[x := c](z) \approx \tau_{\beta_x^{\overline{c}}}(z)$. From Lemma 6.8 it follows that $\varphi[\tau_\beta[x := c]] \approx \varphi[\tau_{\beta_x^{\overline{c}}}]$. Then symmetry, (R11) and Modus Ponens imply the equivalence (**).

Finally, we show that the Bridge Property holds. By definition, $\Gamma(a) = \overline{c}$ iff $c \equiv a \in \Phi$, for any $a \in Expr_\mathcal{L}$. Thus, $\Gamma(a) = \overline{c} \in TRUE \Leftrightarrow c \in \Phi \Leftrightarrow a \in \Phi \Leftrightarrow a \in T = \mathcal{L}(\Phi) = \Phi \cap Expr_\mathcal{L}$. The second part of the Bridge Property follows from the fact that in this classical setting the set $FALSE$ is the complement of $TRUE$ in $M$, and $a \in Expr_\mathcal{L} \smallsetminus T$ iff $\sim a \in T$.

The following equivalences show that $(\mathcal{M}, \vartheta)$ is in fact a model of the maximally $\mathcal{K}_C^*$-consistent Henkin set $\Phi$:

$$(\mathcal{M}, \vartheta) \vDash \varphi \iff \Gamma(\varphi, \vartheta) \in TRUE \stackrel{(6.1)}{\iff} \overline{\varphi[\tau_\vartheta]} \in TRUE \stackrel{(6.3)}{\iff} \varphi \in \Phi.$$

□

**Theorem 6.11** *Every $\mathcal{K}_C^*$-consistent set has a model.*

**Proof.** Suppose $\Phi \subseteq Expr(\mathcal{L}, C)$ is $\mathcal{K}_C^*$-consistent. Let $C^*$ be a set of new constant symbols with $|C^*| = |Expr(\mathcal{L}, C)| =: \kappa$, and let $C' := C \cup C^*$. Then $|Expr(\mathcal{L}, C')| = \kappa$. Let $(\varphi_\alpha \mid \alpha < \kappa)$ be an enumeration of $Expr(\mathcal{L}, C')$. By induction on $\alpha < \kappa$ we construct a chain $(\Phi_\alpha)_{\alpha < \kappa}$ of $\mathcal{K}_{C'}^*$-consistent sets $\Phi_\alpha \subseteq Expr(\mathcal{L}, C')$ containing $\Phi$ such that $\Phi' = \bigcup_{\alpha < \kappa} \Phi_\alpha$ is a maximally $\mathcal{K}_{C'}^*$-consistent Henkin-set. Then we may apply Theorem 6.10.

Put $\Phi_0 := \Phi$. For limit ordinals $\lambda$ we define $\Phi_\lambda := \bigcup\{\Phi_\alpha \mid \alpha < \lambda\}$. If $\alpha = \delta + 1$ is a successor ordinal, then we set

$$\Phi'_\alpha := \begin{cases} \Phi_\delta \cup \{\varphi_\delta\}, & \text{if } \Phi_\delta \cup \{\varphi_\delta\} \text{ is } \mathcal{K}^*_{C'}\text{ - consistent} \\ \Phi_\delta \cup \{\varphi_\delta : false\} & \text{else.} \end{cases}$$

By (R4) and the $\mathcal{K}^*_{C'}$-consistency of $\Phi_\delta$, $\Phi'_\alpha$ is $\mathcal{K}^*_{C'}$-consistent. The set $\Phi'_\alpha$ contains at most $|\alpha|$-many constant symbols from $C^*$. Thus, there is a constant symbol $c \in C^* \setminus con(\Phi'_\alpha)$.[23] Define

$$\Phi_\alpha := \begin{cases} \Phi'_\alpha \cup \{\psi[x := c]\}, & \text{if } \varphi_\delta \in \Phi'_\alpha \text{ and } \varphi_\delta \text{ has the form } \exists x.\psi \\ \Phi'_\alpha, & \text{else.} \end{cases}$$

We show that $\Phi_\alpha$ is $\mathcal{K}^*_{C'}$-consistent. Since $\Phi'_\alpha$ is $\mathcal{K}^*_{C'}$-consistent, it is enough to consider the case $\Phi_\alpha = \Phi'_\alpha \cup \{\psi[x := c]\}$. Note that $\Phi'_\alpha = \Phi_\delta \cup \{\exists x.\psi\}$. Towards a contradiction suppose that $\Phi_\alpha$ is $\mathcal{K}^*_{C'}$-inconsistent. Let $\bot$ be a $\mathcal{K}^*_{C'}$-inconsistent formula (e.g., $(x \equiv x) : false$). Then, by Lemma 6.4, there is a derivation of the sequent

$$\Phi'_\alpha \cup \{\psi[x := c]\} \vdash \bot.$$

Since derivation is finitary, we may assume that $\Phi'_\alpha$ is a finite set and thus contains only finitely many variables. Let $x_c$ be any variable not occurring in the above derivation. By Lemma 6.1, the following is a derivation, too:

$$\Phi'_\alpha[c := x_c] \cup \{\psi[x := c][c := x_c]\} \vdash \bot[c := x_c].$$

We apply the rule (R9) and obtain

$$\Phi'_\alpha \cup \{\exists x.\psi\} \vdash \bot.$$

Since $\exists x.\psi \in \Phi'_\alpha$, we get

$$\Phi'_\alpha \vdash \bot.$$

This is a contradiction to the $\mathcal{K}^*_{C'}$-consistency of $\Phi'_\alpha$. It follows that $\Phi'$ is $\mathcal{K}^*_C$-consistent. By Theorem 6.10, it has a model $(\mathcal{M}', \vartheta)$ with respect to the language $Expr(\mathcal{L}, C')$. Recall that $C \subseteq C'$. If we restrict the Gamma-function $\Gamma' : Expr(\mathcal{L}, C') \times M^V \to M$ of $\mathcal{M}'$ to the function $\Gamma : Expr(\mathcal{L}, C) \times M^V \to M$, then obviously we obtain a model $\mathcal{M}$ with Gamma-function $\Gamma$ such that for all formulas $\varphi \in Expr(\mathcal{L}, C)$ we have $(\mathcal{M}, \vartheta) \vDash \varphi \Leftrightarrow (\mathcal{M}', \vartheta) \vDash \varphi$. Such a model is called the reduct of $\mathcal{M}'$ to the sublanguage $Expr(\mathcal{L}, C)$ (see [25, 29]). In particular, $(\mathcal{M}, \vartheta) \vDash \Phi$. $\square$

---

[23]$con(\Phi'_\alpha)$ denotes the set of all constant symbols occurring in $\Phi'_\alpha$.

**Theorem 6.12** *For every classical abstract logic $\mathcal{L}$ and every set of constant symbols $C$, the calculus $\mathcal{K}_C^*$ is complete. That is, for every set $\Phi \cup \{\varphi\} \subseteq Expr(\mathcal{L}, C)$ we have*

$$\Phi \Vdash \varphi \Rightarrow \Phi \vdash \varphi.$$

**Proof.** Suppose $\Phi \nvdash \varphi$. Then $\Phi \cup \{\varphi : false\}$ is $\mathcal{K}_C^*$-consistent (otherwise, the rules (R1) and (R2) would yield a derivation of $\Phi \vdash \varphi$). By Theorem 6.11, the set $\Phi \cup \{\varphi : false\}$ has a model. That is, $\Phi \nVdash \varphi$. $\square$

# References

[1] J.-Y. Beziau, *Bivalent Semantics for De Morgan Logic (The Uselessness of Four-valuedness)*, in: W.A.Carnielli, M.E.Coniglio and I.M.L.D'Ottaviano (eds), *The many sides of logic*, College Publication, London, 391 – 402, 2009.

[2] D.J. Brown and R. Suszko, *Abstract Logics*, Dissertationes Mathematicae 102, 9 – 42, 1973.

[3] S.L. Bloom and D.J. Brown, *Classical Abstract Logics*, Dissertationes Mathematicae 102, 43 – 51, 1973.

[4] S. L. Bloom and R. Suszko, *Investigation into the sentential calculus with identity*, Notre Dame Journal of Formal Logic 13(3), 289 – 308, 1972.

[5] B.A. Davey and H.A. Priestley, *Introduction to Lattices and Order*, 2. ed., Cambridge University Press, 2002.

[6] J.M. Dunn, G.H. Hardegree, *Algebraic Methods in Philosophical Logic*, Clarendon Press, Oxford, 2001.

[7] J. M. Font, *Belnap's Four-Valued Logic and De Morgan Lattices*, Logic Journal of the IGPL 5(3), 1 – 29, 1997.

[8] J.M. Font, R. Jansana and D. Pigozzi, *A survey of abstract algebraic logic*, Studia Logica 74, 13 –97, 2003.

[9] J.M. Font and V. Verdú, *A first approach to abstract modal logics*, Journal of Symbolic Logic 54(3), 1042 – 1062, 1989.

[10] H. Hermes, *Term Logic with Choice Operator*, Springer Verlag, 1970.

[11] S. Lewitzka, $\in_T$ ($\Sigma$)-*Logik: Eine Erweiterung der Prädikatenlogik erster Stufe mit Selbstreferenz und totalem Wahrheitsprädikat*, Diplomarbeit, Technische Universität Berlin, 1998.

[12] S. Lewitzka, *Abstract Logics, Logic Maps and Logic Homomorphisms*, Logica Universalis 1(2), 243 – 276, 2007.

[13] S. Lewitzka, $\in_4$: *a 4-valued truth theory and meta-logic*, manuscript, 2007.

[14] S. Lewitzka, *A sequent calculus for $\in_4$-Logic*, manuscript, 2007.

[15] S. Lewitzka, $\in_I$: *an intuitioninistic logic without Fregean axiom and with predicates for truth and falsity*, Notre Dame Journal of Formal Logic 50(3), 275 – 301, 2009.

[16] S. Lewitzka, $\in_K$: *a non-Fregean logic of explicit knowledge*, Studia Logica 97(2), 233 – 264, 2011.

[17] S. Lewitzka, *Semantically closed intuitionistic abstract logics*, Journal of Logic and Computation 22(3), 351 – 374, 2012

[18] S. Lewitzka, *Construction of a canonical model for a first-order non-Fregean logic with a connective for reference and a total truth predicate*, The Logic Journal of the IGPL, DOI: 10.1093/jigpal/jzr050, 2012.

[19] S. Lewitzka, *Necessity as justified truth*, arXiv:1205.3803v2, 2012.

[20] S. Lewitzka and A.B.M. Brunner, *Minimally generated abstract logics*, Logica Universalis 3(2), 219 – 241, 2000.

[21] Y. Ma, G. Qi and P. Hitzler, *Computing inconsistency measure based on paraconsistent semantics*, Journal of Logic and Computation 21 (6), 1257 – 1281, 2011.

[22] G. Priest, *An Introduction to Non-Classical Logic*, Cambridge University Press, 2001.

[23] H. Rasiowa, *An Algebraic Approach to Non-Classical Logic*, North-Holland Publ. Co., Amsterdam, 1974.

[24] K. Robering, *Logics with Propositional Quantifiers and Propositional Identity*, in: S. Bab, K. Robering (eds.), *Judgements and Propositions*, Logos Verlag, Berlin, 2010.

[25] W. Sträter, $\in_T$ Eine Logik erster Stufe mit Selbstreferenz und totalem Wahrheitsprädikat, Dissertation, KIT-Report 98, Technische Universität Berlin, 1992.

[26] R. Suszko, *Non-Fregean Logic and Theories*, Analele Universitatii Bucuresti, Acta Logica 11, 105 – 125, 1968.

[27] R. Suszko, *Abolition of the Fregean Axiom*, in: R. Parikh (ed.) *Logic Colloquium*, Lecture Notes in Mathematics 453, 169 – 239, Springer Verlag 1975.

[28] B. C. van Fraassen, *Formal Semantics and Logic*, The Macmillan Company, New York, 1971.

[29] Ph. Zeitz, *Parametrisierte $\in_T$-Logik – eine Theorie der Erweiterung abstrakter Logiken um die Konzepte Wahrheit, Referenz und klassische Negation*, Dissertation, Logos Verlag Berlin, 2000.